\documentclass[aps,prd,showpacs,superscriptaddress,amsmath,amsfont,amssymb,preprint,nofootinbib]{revtex4}
\pdfoutput=1
\usepackage{color,graphicx,epsfig}
\usepackage{ifpdf}
\usepackage{amsmath}
\usepackage{bm}
\usepackage{color}
\usepackage[english]{babel}
\usepackage{graphicx}
\usepackage{amsfonts}
\usepackage{amssymb}
\usepackage{hyperref}
\usepackage{enumerate}

\definecolor{nicered}{rgb}{0.7,0.1,0.1}
\definecolor{nicegreen}{rgb}{0.1,0.5,0.1}
\hypersetup{colorlinks,citecolor= nicegreen,linkcolor= nicered}

\newcommand{\bea}{\begin{eqnarray}}
\newcommand{\eea}{\end{eqnarray}}

\definecolor{Red}{rgb}{1.,0.,0.}

\def\gsim{{~\raise.15em\hbox{$>$}\kern-.85em
          \lower.35em\hbox{$\sim$}~}}
\def\lsim{{~\raise.15em\hbox{$<$}\kern-.85em
          \lower.35em\hbox{$\sim$}~}}

\newcommand{\Rrd}{{\mathcal R}}

\newcommand{\es}[2] {\begin{equation} \label{#1} \begin{split} #2 \end{split} \end{equation}}

\definecolor{nicered}{rgb}{0.7,0.1,0.1}
\definecolor{nicegreen}{rgb}{0.1,0.5,0.1}
\hypersetup{colorlinks,citecolor= nicegreen,linkcolor= nicered}

\newcommand{\beq}{\begin{equation}}
\newcommand{\eeq}{\end{equation}}
\newcommand{\tr}{\mathrm{Tr}}

\newcommand{\cL}{\mathcal{L}}

\newcommand{\BR}{{\rm BR}}

\newcommand{\Rprod}{R_{13/8}}

\def\mysection#1{{{\bf #1}.~}}
\def\OMIT#1{}

\begin{document}

\def\Cincy{Department of Physics, University of Cincinnati, Cincinnati, Ohio 45221, USA}
\def\MIT{Center for Theoretical Physics, Massachusetts Institute of Technology, Cambridge, MA 02139, USA}
\def\Annecy{LAPTh, Universit\' e Savoie Mont Blanc, CNRS B.P. 110, F-74941 Annecy-le-Vieux, France}
\def\CERN{Theory Division, CERN, 1211 Geneva 23, Switzerland}
\def\JSI{Jozef Stefan Institute, Jamova 39, 100 Ljuljana, Slovenia}
\def\UL{Faculty of Mathematics and Physics, University of Ljubljana, Jadranska 19, 1000 Ljubljana, Slovenia}

\preprint{MIT-CTP/4786}

\title{
Comments on the diphoton excess:\\
critical reappraisal of effective field theory interpretations 
}

\author{Jernej F. Kamenik} 
\email[Electronic address:]{jernej.kamenik@cern.ch} 
\affiliation{\JSI}
\affiliation{\UL}

\author{Benjamin R.  Safdi} 
\email[Electronic address:]{bsafdi@mit.edu} 
\affiliation{\MIT}

\author{Yotam Soreq} 
\email[Electronic address:]{soreqy@mit.edu} 
\affiliation{\MIT}

\author{Jure Zupan} 
\email[Electronic address:]{zupanje@ucmail.uc.edu} 
\affiliation{\Cincy}

\begin{abstract}
We consider the diphoton excess observed by ATLAS and CMS using the most up-to-date data and estimate the preferred enhancement in the production rate between 8 TeV and 13 TeV.
Within the framework of effective field theory~(EFT), we then show that 
for both spin-$0$ and spin-$2$ Standard Model~(SM) gauge-singlet resonances, two of the three processes $S\to ZZ$, $S\to Z\gamma$,  and $S\to WW$ must occur with a non-zero rate.  Moreover, we demonstrate that these branching ratios are highly correlated in the EFT. 
Couplings of $S$ to additional SM states may be constrained and differentiated 
by comparing the $S$ production rates with and without the vector-boson fusion~(VBF) cuts. 
We find that for a given VBF to inclusive production ratio there is maximum rate of $S$ to gauge bosons, $b \bar b$, and lighter quark anti-quark pairs.  Simultaneous measurements of the width and the VBF ratio may be able to point towards the existence of hidden decays.
\end{abstract}

\maketitle
\section{Introduction} \label{sec:introduction}

First data at the 13\,TeV LHC hint to the existence of a diphoton resonance, $S$, with a mass $m_S\sim$750 GeV~\cite{ATLAS-CONF-2015-081,ATLAS:Moriond,CMS:2015dxe,CMS:Moriond, CMS:2016owr}.  Because $S$ decays to two photons, $S\to \gamma\gamma$, one generically expects that $S$ also decays to the other gauge boson pairs: $S\to ZZ$, $S\to Z\gamma$,  and $S\to WW$. The aim of this manuscript is to make this generic expectation more precise and to consider implications of couplings to other Standard Model (SM) states.

We begin our discussion by performing an up-to-date fit of $S$ to the available data (after Moriond EW~2016). We quantify the enhancement of the production cross section between the 8\,TeV and 13\,TeV LHC runs implied by the current data. The large preferred enhancement can be interpreted as a  preference for heavy quark annihilation or gluon fusion dominated production.\footnote{For alternative production mechanisms leading to large ratios of 13\,TeV and 8\,TeV production cross-sections c.f.~\cite{Franceschini:2015kwy, Knapen:2015dap,Liu:2015yec,Huang:2015evq}. } Accordingly, we extract the preferred diphoton signal strengths for various possible production modes. 

One of our central results is that, within our working assumptions, at least two out of three branching ratios, $S\to ZZ$, $S\to Z\gamma$,  and $S\to WW$, need to be nonzero if the resonance is a SM gauge singlet.\footnote{We also show that the branching ratios are constrained if $S$ is part of an $SU(2)_L$ doublet or triplet representation; in general, we find that in the EFT framework at least one of the additional branching ratios to electroweak gauge bosons needs to be non vanishing.}
Our working assumptions are: (1)~that the resonance $S$ is either a spin-0 or spin-2 particle, and (2)~that effective field theory~(EFT) may be used to describe the interactions of $S$ with the SM. Implicitly, this means that we can truncate the EFT after the first few lowest dimension operators. In our case, we keep systematically all the terms up to and including operators of dimension 5 (dimension 6 for $S$ that is an electroweak doublet). 

If the mixing of $S$ with the SM-Higgs and its coupling with the Higgs kinetic mixing can be neglected, two branching ratios (e.g., $S\to ZZ$ and $S\to WW$) are predicted in terms of the third one (in this case, $S\to Z\gamma$). If $S$ mixes with the Higgs, then measuring two branching ratios out of three, $S\to ZZ$, $S\to Z\gamma$,  or $S\to WW$, predicts the third. Below, we derive the sum rules relating these branching ratios. While the importance of these decay modes has already been stressed in the literature~\cite{Franceschini:2015kwy,Low:2015qep,Low:2015qho,Alves:2015jgx,Altmannshofer:2015xfo,DiChiara:2015vdm,Ellis:2015oso,Petersson:2015mkr,Cao:2015pto,Kobakhidze:2015ldh,Fichet:2015vvy,Han:2015dlp,Feng:2015wil,Heckman:2015kqk,Berthier:2015vbb,Craig:2015lra,Dev:2015vjd,Stolarski:2016dpa,Fichet:2016pvq,Buttazzo:2015txu,Gupta:2015zzs,Agrawal:2015dbf}, we phrase the discussion directly in terms of observables, making contact with the experiments very explicit. 

Determining the dominant production channel(s) of the resonance 
 is paramount as more data accumulates.  Towards that end, we consider the simultaneous measurements of the rate after applying vector boson fusion~(VBF) cuts along with the total width. This can distinguish between different production channels and help resolve  whether hidden decays are required.  For example, we show that the ratio of the rates with and without applying VBF cuts is an efficient discriminator between gluon fusion and heavy quark production.  Furthermore, for a given VBF ratio there is a maximum allowed rate to electroweak gauge bosons and quark pairs (excluding $t \bar t$); measuring a rate beyond this value may indicate decays to a hidden sector or to currently relatively unconstrained final states, such as $t \bar t$.

The paper is organized as follows.  In Section~\ref{sec:fit} we present an updated fit of the resonance to the most recent ATLAS and CMS 13\,TeV and 8\,TeV analyses. In Sec.~\ref{sec:EFT} we introduce the EFT interactions of $S$  to the SM particles, assuming $S$ is either spin-$0$ and spin-$2$.  In Sec.~\ref{sec:correlations} we derive the correlations between branching ratios of $S$ to different EW gauge boson pairs.  Section \ref{sec:VBF} discusses the importance of searching for the potential VBF production of $S$, while Sec.~\ref{sec:SMrate} considers the implications of simultaneous measurements of the VBF production rate and the total width.  We conclude in Sec.~\ref{sec:Conclusions}. 

\section{Fit to the current data} \label{sec:fit}

\begin{figure}[!t]
\centering
\includegraphics[width=0.45\textwidth]{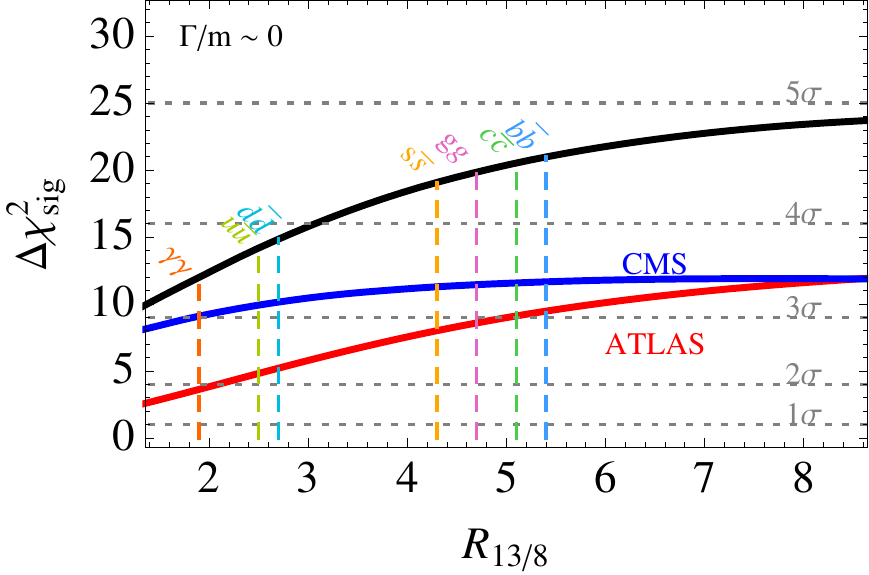}~~~
\includegraphics[width=0.45\textwidth]{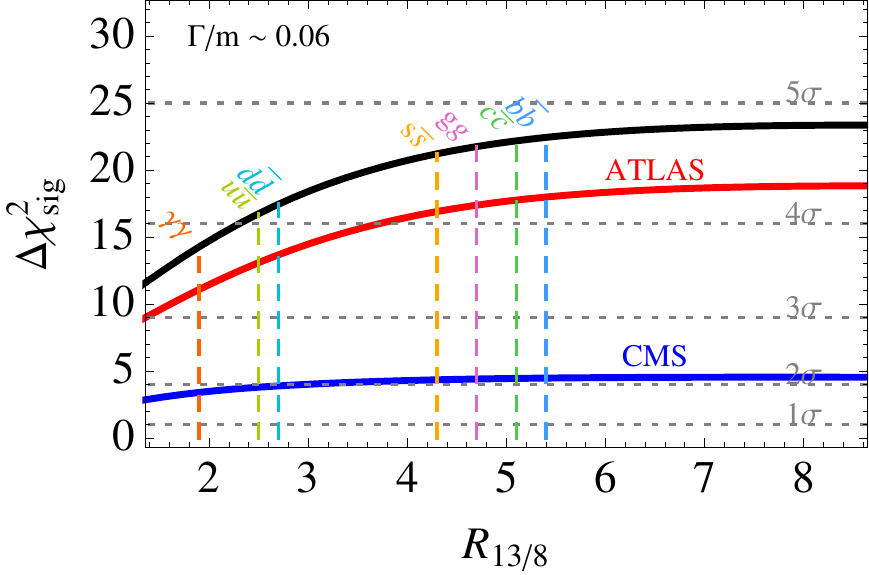}\\
\caption{The significance of the diphoton excess as a function of $\Rprod$, the ratio of the 13\,TeV and 8\,TeV production cross sections, for narrow (left panel) and wide (right panel) width hypotheses. The blue, red and black curves correspond to CMS only data, ATLAS only data and the combined fit, respectively. The vertical lines indicate different production mechanisms computed using the NLO NNPDF 2.3~\cite{Ball:2014uwa} pdf set.
}
\label{fig:FitSignificant}
\end{figure}

\begin{figure}[!t]
\centering
\includegraphics[width=0.45\textwidth]{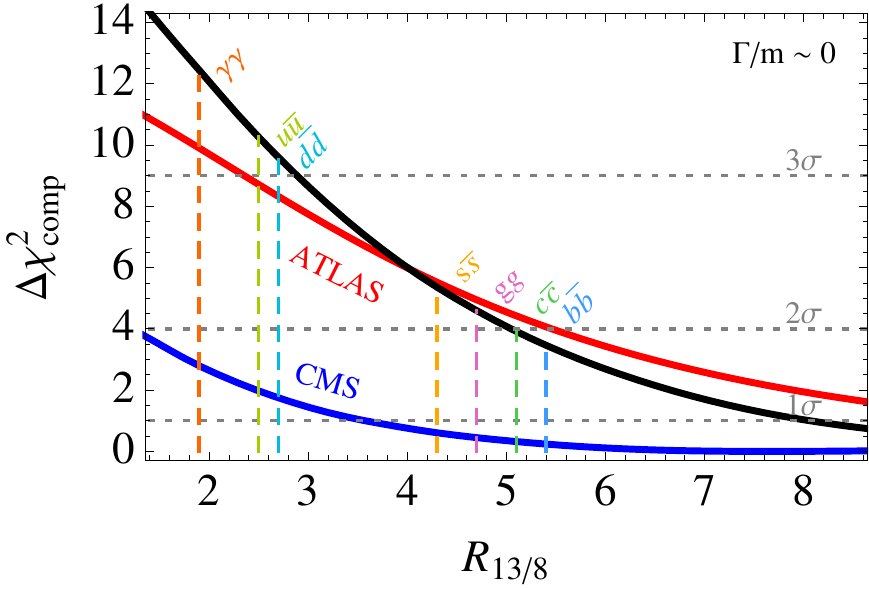}~~~
\includegraphics[width=0.45\textwidth]{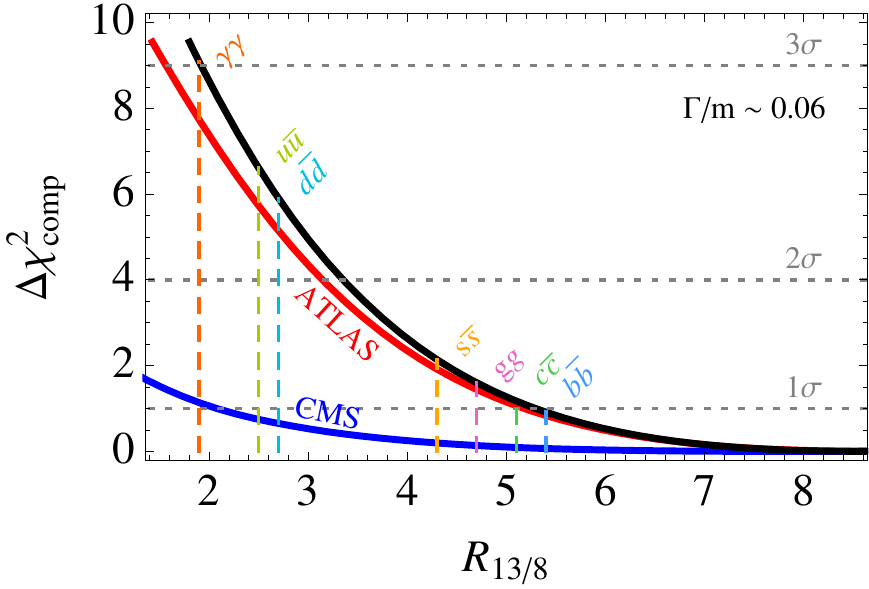}\\
\caption{The compatibility between the 8\,TeV and the 13\,TeV data sets as a function of the ratio between the 8\,TeV and 13\,TeV production, $\Rprod$. The blue, red and black curves correspond to only CMS, only ATLAS and combined fit. The vertical lines indicate different production mechanisms  (computed using NLO NNPDF 2.3~\cite{Ball:2014uwa} pdf set). Left: for narrow width hypothesis; Right: for wide width hypothesis.
}
\label{fig:FitCompatability}
\end{figure}

We start our discussion of the diphoton excess by performing a $\chi^2$ analysis of the current publicly available  ATLAS and CMS data.  Our analysis addresses the following questions: 
(i)~What is the significance of the excess after combining all publicly available ATLAS and CMS data? 
(ii)~What is the preferred production channel for $S$?
(iii)~What is the compatibility between the 8\,TeV and 13\,TeV data sets? 
(iv)~What is the preferred diphoton production cross-section, $\sigma_{\gamma\gamma} \equiv \sigma(pp\to X)_{13}\BR_{\gamma\gamma}$, at 13\, TeV?  In the combination below we resort to several approximations. Most importantly, we assume  constant efficiencies for different channels. The results should thus be taken as a rough guide only.
 
The production mechanism of $S$ is presently unknown. Some handle on it can be obtained already now, though, by comparing the LHC data collected at the 8\,TeV and 13\,TeV center of mass energies. For this purpose, we examine the dependence of the current experimental results on the ratio between the 8\,TeV and the 13\,TeV $S$ production rates
\begin{align}
	\Rprod \equiv \frac{\sigma(pp\to S)_{13}}{\sigma(pp\to S)_{8}} \, .
\end{align}
 Examining the compatibility between the two sets of measurements gives valuable information on the production mechanism because different parton luminosities scale differently with collider energy. 

In our analysis we include the 8\,TeV and 13\,TeV diphoton searches, including the Moriond EW 2016 updates, by ATLAS~\cite{ATLAS:Moriond} and CMS~\cite{CMS:Moriond, CMS:2016owr} and distinguish between the narrow and wide decay width hypotheses.
In combining the results we assume uncorrelated measurements  and construct a $\chi^2$ as function of $\Rprod$.
  For CMS  and  narrow-width approximation we use the reported $\chi^2$ functions, $\chi^2_{\rm CMS, 8/13}$ (see Fig.~10 of~\cite{CMS:2016owr}),
\begin{align}\label{eq:CMS:chi2}
	\chi^2_{\rm CMS}(\sigma_{\gamma\gamma},\,\Rprod) 
=  	\chi^2_{\rm CMS,8}\left(\sigma_{\gamma\gamma} \cdot \Rprod^{\rm ref}/\Rprod \right) + \chi^2_{\rm CMS,13}(\sigma_{\gamma\gamma}), 
\end{align}
where $\Rprod^{\rm ref}=4.7\,(4.2)$ was chosen as the reference value by CMS for the spin-0\,(spin-2) hypothesis.
Here $\sigma_{\gamma\gamma}$
is the diphoton signal rate at 13 TeV.
For the wide resonance hypothesis, $\Gamma_S/m_S \sim 0.06$, CMS does not provide the $\chi^2$ functions directly. 
However, we can construct a quadratic $\chi^2$ functions based on the public $p$-value distributions for  8\,TeV and 13\,TeV.  These are described by two parameters each, the two minima and the two curvatures.

Similarly, for ATLAS results the $\chi^2_{\rm ATLAS}(\sigma_{\gamma\gamma},\,\Rprod) $ is defined analogously to \eqref{eq:CMS:chi2}, but assuming quadratic functions for $\chi^2_{\rm ATLAS,8}, \chi^2_{\rm ATLAS,13}$. In this case, the parameters are fixed using 
the quoted significance of the excess for the 13 TeV and 8 TeV analyses, the compatibility of the two, and the global minimum $\chi_{\gamma\gamma}$. The first three inputs are provided in ~\cite{ATLAS:Moriond} except for the narrow width case where the significance of the $8$\,TeV data (re)analysis is not provided. We estimate this by fitting the signal to a single $m_{\gamma\gamma}$ bin.  Such a narrow resonance fit to a binned distribution might not faithfully represent the maximal significance of the constraint.  In order to be conservative, we also consider the three bins nearest to $m_S$ and employ the weakest constraint in the combination fit.

The global minimum $\chi_{\gamma\gamma}$ we obtained from  
\begin{align}\label{eq:ATLAS:chi2}
	\chi^2_{\rm ATLAS, approx.}(\sigma_{\gamma\gamma},\,\Rprod)  &
=	\sum_{i\in 8\,{\rm TeV}} \frac{(N_i - N_{i,{\rm bkg}} - \sigma_{\gamma\gamma} \cL_{8} \mathcal R_i \epsilon_i/\Rprod)^2}{\sigma^2_i} \nonumber\\
&	+\sum_{i\in 13\,{\rm TeV}} \frac{(N_i - N_{i,{\rm bkg}} - \sigma_{\gamma\gamma} \cL_{13}  \mathcal R_i \epsilon_i)^2}{\sigma^2_i} \, ,
\end{align}
where $\cL_{8(13)}=20.3 (3.2)~{\rm fb}^{-1}$ are the 8\,TeV (13\,TeV) integrated luminosities 
and $i$ runs over the relevant  data points presented in~\cite{ATLAS:Moriond}. The $N_i (N_{i,{\rm bkg}})$ are the observed (estimated background) number of events in the $i$-th bin and $\sigma_i$ the corresponding estimated uncertainty.
Signal is modeled using a normalized Breit-Wigner resonance function centered at $m_S$ with a width $\Gamma_S$, whose integral over the $i$-th bin is given by  $\mathcal R_i$. Finally, $\epsilon_i$ is the corresponding signal efficiency for the 8\,(13)\,TeV analysis. ATLAS presented two analyses with different cuts  on the transverse energies of the photons. In the following, we employ the ``spin-0" analysis which requires $p_T(\gamma_{1(2)})>0.4\,(0.3) m_{\gamma\gamma}$. While it contains only a subset of data passing the cuts of the ``spin-2" analysis, it exhibits a slightly more significant excess and reduced tension with the 8\,TeV results. Using MadGraph 5~\cite{Alwall:2014hca} simulations we estimate the signal efficiency to be $\epsilon_i \simeq 0.65\,(0.40)$ close to $m_{\gamma\gamma}\sim m_S$ for a spin-0\,(spin-2) resonance $S$. The obtained spin-0 efficiency is consistent with the range quoted in~\cite{ATLAS:Moriond}. 

Next, the ATLAS and CMS $\chi^2$ functions  are marginalized over $\sigma_{\gamma\gamma}$
 \begin{align}
	\chi^2_{\rm ATLAS/CMS} (\Rprod) = {\rm min}_{\sigma_{\gamma\gamma}} \left\{ \chi^2_{\rm ATLAS/CMS}(\sigma_{\gamma\gamma},\,\Rprod) \right\} \, .
\end{align}
Comparing the marginalized $\chi^2$ with the zero-signal hypothesis,
\begin{align}
	\Delta \chi^2_{\rm sig} = {\chi^2_{} (\Rprod) - \chi^2(0,\Rprod)} \,,
\end{align} 
gives the significance of the excess.
We plot $\Delta \chi^2_{\rm sig}$ in Fig.~\ref{fig:FitSignificant}  assuming a spin-0 $S$ with either a narrow (left panel) or wide decay width (right panel). At present the difference between spin-0 and spin-2 hypotheses is negligible, so that we do not plot the corresponding results for spin-2.  The vertical lines  in Fig.~\ref{fig:FitSignificant}  indicate $\Rprod$ ratios expected for different production mechanisms, computed using the NLO NNPDF 2.3~\cite{Ball:2014uwa} pdf set. The excess becomes more significant for larger values of $\Rprod$, i.e., for larger ratios of 13\,TeV to 8\,TeV production cross sections.

This feature can be seen also from the compatibility of the 8\,TeV and 13\,TeV datasets, which can be assessed through 
\begin{align}
	\Delta \chi^2_{\rm comp} = { \chi^2(\Rprod) - \chi^2_{\rm min}} \,.
\end{align} 
 Here $\chi^2_{\rm min}$ is the minimum of $\chi^2(\Rprod) $  when varying $\Rprod$. The dependence of $\Delta \chi^2_{\rm comp}$ on $\Rprod$ is shown in Fig.~\ref{fig:FitCompatability}.
 Since the significance of the excess in the 8\,TeV data is much smaller than for the 13\,TeV measurements, the fit prefers a large enhancement of the 13\,TeV production rates compared to 8\,TeV. The above may be interpreted as a preference for  having sea partons (gluons or heavy quarks) as initial states. These predict higher $\Rprod$ ratios, $\Rprod\gtrsim 5$. In contrast, $\Rprod \lesssim 3(4)$ if $S$ is produced through valence quark annihilation or photon fusion~\cite{Fichet:2015vvy, Csaki:2016raa, Fichet:2016pvq, Harland-Lang:2016apc, Ababekri:2016kkj, Harland-Lang:2016qjy} which is disfavored by more than $3\,(2)\,\sigma$ for a narrow\,(wide) $S$. In the following we will consider these results as indicative but keep the possibility of valence quark annihilation and photon (or more generally EW vector boson) fusion dominated $S$ production open. The results for spin-0 and spin-2 are, again, very similar. 

\begin{figure}[!t]
\centering
\includegraphics[width=0.45\textwidth]{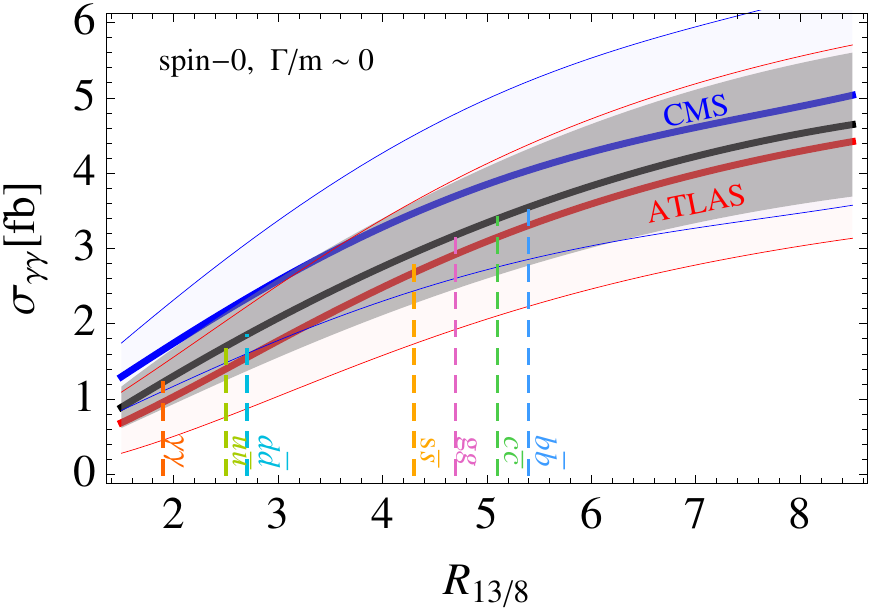}~~~
\includegraphics[width=0.45\textwidth]{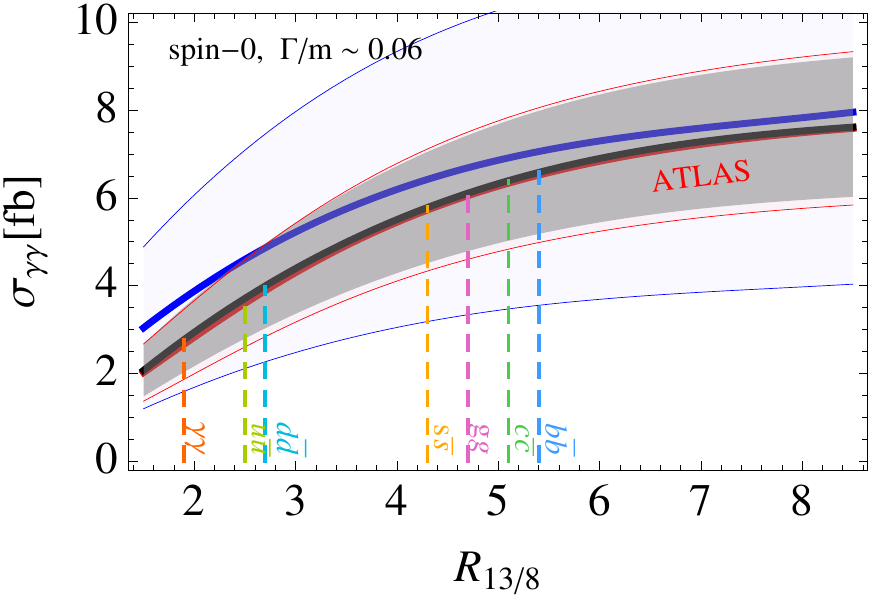}\\
\includegraphics[width=0.45\textwidth]{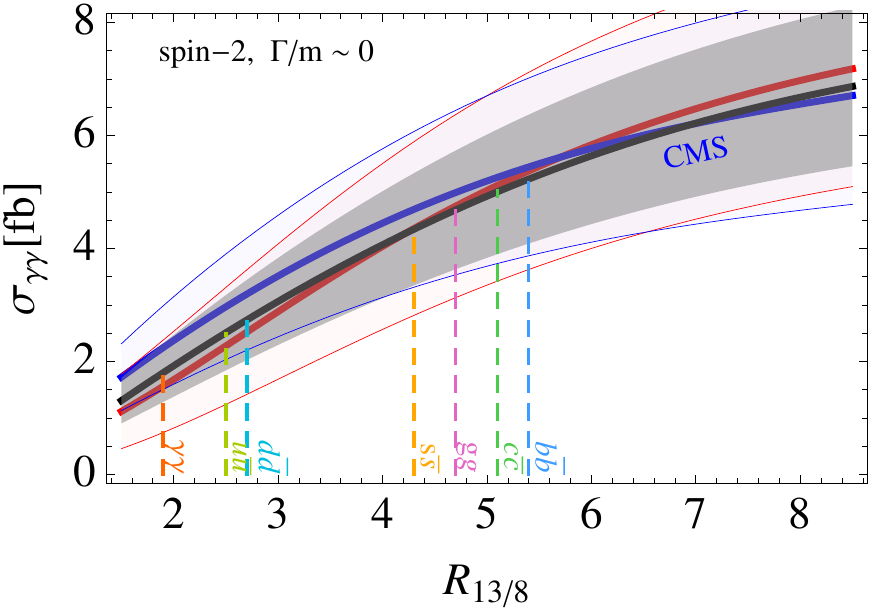}~~~
\includegraphics[width=0.45\textwidth]{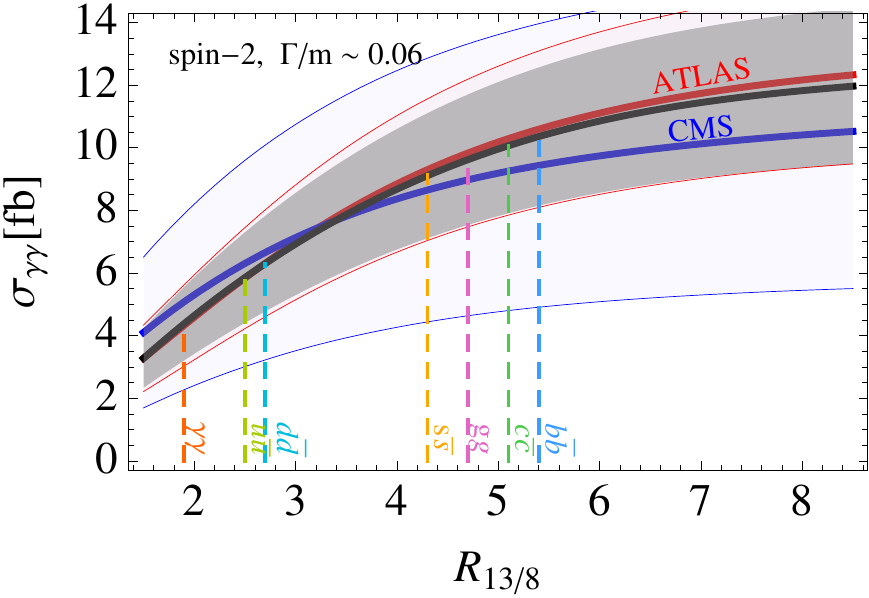}
\caption{ The preferred 13 TeV diphoton rate, $\sigma_{\gamma\gamma}$, as a function of the 13\,TeV to 8\,TeV production cross section ratio, $\Rprod$. The blue, red, and black solid lines give the CMS only, ATLAS only and combined fit results, respectively, with 1$\sigma$ bends denoted by blue, red thin lines and the grey band. The vertical lines indicate different production mechanisms computed using NLO NNPDF 2.3~\cite{Ball:2014uwa} pdf set. Left (Right): for narrow (wide) width hypothesis; Up (Down): for spin-0\,(2). 
}
\label{fig:Xsec}
\end{figure}

Finally, we combine the ATLAS and CMS data to estimate the preferred value for the cross section $\sigma_{\gamma\gamma}$. We use the combined $\chi^2$ \eqref{eq:CMS:chi2}, \eqref{eq:ATLAS:chi2} to find the $1\sigma$ band as function of $\Rprod$. The results are presented in Fig.~\ref{fig:Xsec}. Note that the best-fit cross section grows larger as the ratio $\Rprod$ increases.  
In Tab.~\ref{tab:Xsec} we summarize the best-fit 13 TeV diphoton rates, $\sigma_{\gamma\gamma}$, for a number of assumed production mechanisms. In Tab.~\ref{tab:Xsec} we take the efficiencies, $\epsilon_i$, to be independent of the production mechanism.
The errors due to this approximation are expected to be subleading compared to the current sizable experimental uncertainties and the limitations of our fitting procedure. For the spin-0 case one can understand the smallness of these effects most easily by noting that the photon $p_T$ distributions are boost invariant and thus at parton level independent of the particular parton luminosity integration.\footnote{We thank Gilad Perez for insightful discussions on this point.}

In the subsequent sections we will also make use of experimental constraints on the branching ratios of $S$ to different final states, such as $ZZ$, $WW$, $Zh$, $hh$ and $f\bar f$.  These constraints are taken from Tab.~1 of~\cite{Franceschini:2015kwy} (see also~\cite{Gupta:2015zzs}). Using the 8\,TeV data~\cite{Aad:2015kna,ATLAS:2014rxa,Khachatryan:2015cwa,Aad:2015agg} one has at 95\,\%~CL
\begin{align}
	\label{eq:RWWbound}
	\Rrd_{WW} &\lesssim 40\left(\frac{\Rprod\, 5\rm fb}{5 \sigma_{\gamma\gamma}}\right) \approx \frac{110\rm \, fb}{\sigma_{\gamma\gamma}} \, , \\
	\Rrd_{ZZ} &\lesssim 12 \left(\frac{\Rprod\, 5\rm fb}{5 \sigma_{\gamma\gamma}}\right) \approx  \frac{65\rm \,fb}{\sigma_{\gamma\gamma}}\, , \\ 
	\Rrd_{Zh} &\lesssim 20 \left(\frac{\Rprod\, 5\rm fb}{5 \sigma_{\gamma\gamma}}\right) \approx  \frac{110\rm \,fb}{\sigma_{\gamma\gamma}}\, , \\ 
	\label{eq:Rhhbound}
	\Rrd_{hh} &\lesssim 40 \left(\frac{\Rprod \, 5 \rm fb}{5 \sigma_{\gamma\gamma}}\right) \approx  \frac{220\rm \,fb}{\sigma_{\gamma\gamma}} \, ,
\end{align}
 where we have defined 
 \beq \label{eq:Ri:definition}
 \Rrd_f \equiv  \BR_f {\big /} \BR_{\gamma\gamma},
 \eeq
 and in the second step conservatively assumed $\Rprod=5.4$ (as in the case of $b\bar{b}$) for maximal enhancement of the prompt $S$ production between 8\,TeV and 13\,TeV.  The corresponding 13 TeV searches are less sensitive \cite{CMS-PAS-HIG-16-001}. 
For the $Z\gamma$ final state we use the recent 13 TeV bound presented by ATLAS~\cite{ATLAS:zgamma}:
\begin{align}
	\label{eq:RZabound}
	\sigma(pp\to X)_{13}\BR_{Z\gamma} < 28\,{\rm fb} \, 
	\quad \Rightarrow \quad
	\Rrd_{Z\gamma} \lesssim \frac{28\,{\rm fb}}{\sigma_{\gamma\gamma}}~{\rm at ~95\,\%~CL}\,,
\end{align} 
to be compared with the rescaled 8 TeV bound $\Rrd_{Z\gamma} \lesssim 12 \left(\Rprod\, 5\rm \, fb / 5 \sigma_{\gamma\gamma}\right) \approx 65 {\rm fb}/\sigma_{\gamma\gamma}$~\cite{Aad:2014fha}.  The 13 TeV bound in \eqref{eq:RZabound} is $\Rrd_{Z\gamma} \lesssim 4.2$, using the central value $\sigma_{\gamma\gamma}=6.6$ fb for spin-0 $S$ with wide decay width produced from the $b\bar b$ initial state, cf. Table \ref{tab:Xsec}. 

\begin{table}[!t]
\begin{center}
\begin{tabular}{ccccc}
\hline\hline
 & \multicolumn{2}{c}{narrow width ($\Gamma/m\to0$)~~~~~}   & \multicolumn{2}{c}{wide width ($\Gamma/m=6\,$\%)~~~~~}  \\
production mech. ~~& spin-0 & spin-2& spin-0 & spin-2\\
\hline
$\gamma\gamma$ &$1.2\pm0.4$&$1.8\pm0.5$&$2.8\pm0.7$&$4.4\pm1.2$ \\
$gg$ 		&$3.2\pm0.7$&$4.7\pm1.0$&$6.1\pm1.3$&$9.6\pm2.1$\\
$u\bar{u}$ 	&$1.7\pm0.4$&$2.5\pm0.7$&$3.7\pm0.9$&$5.9\pm1.4$\\
$d\bar{d}$ 	&$1.9\pm0.5$&$2.7\pm0.7$&$4.0\pm1.0$&$6.3\pm1.5$\\
$s\bar{s}$ 	&$2.9\pm0.7$&$4.4\pm1.0$&$5.8\pm1.3$&$9.1\pm2.0$\\
$c\bar{c}$ 	&$3.4\pm0.8$&$5.0\pm1.1$&$6.4\pm1.3$&$10.1\pm2.1$\\
$b\bar{b}$ 	&$3.6\pm0.8$&$5.3\pm1.1$&$6.6\pm1.4$&$10.3\pm2.2$\\
\hline\hline
\end{tabular}
\end{center}
\caption{The best fit values of the 13 TeV diphoton rate, $\sigma_{\gamma\gamma}=\sigma(pp\to X)_{13}\BR_{\gamma\gamma}$, in fb for various production mechanisms listed in the first column (see also Fig.~\ref{fig:Xsec}). The relevant values of $\Rprod$ have been computed in~\cite{Franceschini:2015kwy} using the  NLO NNPDF 2.3~\cite{Ball:2014uwa} pdf set.}
\label{tab:Xsec}
\end{table}

\section{Effective Field Theory framework} \label{sec:EFT}

We setup an EFT description of the interactions between $S$ and the SM fields. Our discussion partially overlaps with and extends previous results presented in~\cite{Franceschini:2015kwy,Low:2015qep,Low:2015qho,Alves:2015jgx,Altmannshofer:2015xfo,Ellis:2015oso,Petersson:2015mkr,Cao:2015pto,Kobakhidze:2015ldh,Fichet:2015vvy,Han:2015dlp,Feng:2015wil,Heckman:2015kqk,Berthier:2015vbb,Craig:2015lra,Stolarski:2016dpa,Fichet:2016pvq,Buttazzo:2015txu}. We make two choices for the spin of $S$. We start with the spin-$0$ scenario and assume that $S$ is either an $SU(2)_L$ singlet or triplet, commenting also on the possibility that $S$ is an electroweak doublet. The resulting phenomenology is quite similar also in the case where $S$ has spin-$2$, which we cover next. In all cases we consistently keep all the terms up to and including  dimension 5 (dimension 6 for the doublet) and comment on effects at higher powers.

\subsection{Spin-$0$, $SU(2)_L$ singlet}
\label{sec:spin0singlet}

We first consider the case  where $S$ is a gauge singlet spin-$0$ particle. Assuming CP invariance, the remaining choice is whether 
$S$ is a scalar or a pseudo-scalar.  At the level of observables the differences between these two scenarios are small.  We begin with the scalar case and later mention how the pseudo-scalar case differs.

 The only renormalizable interactions of the scalar $S$ with the SM are through the Higgs doublet field $H$,
\beq\label{eq:dim4Lagr}
{\cal L}_{\rm int}^{(4)}=- \mu_S S H^\dagger H-\frac{\lambda_S}{2} S^2 H^\dagger H.
\eeq
where in the unitary gauge $H=(0,v+h)/\sqrt2$, with $v=246$~GeV.
We will see below that the dimensionful parameter $\mu_S$ is required to be small, $\mu_S\lesssim v$, in order not to induce too large of a mixing between $S$ and the Higgs, $h$. The scalar potential for $S$ contains,  in addition to terms in \eqref{eq:dim4Lagr}, the terms involving only $S$, $\lambda_1 m_S^3 S+(m_S^2/2) S^2 +\lambda_3 m_S S^3+ \lambda_4 S^4$.
For simplicity, we assume that the scalar potential for $S$ does not introduce a vacuum expectation value for $S$.   If this is not the case, one can simply shift $S\to S-\langle S \rangle$, and then appropriately redefine the coefficients in the SM and the interaction Lagrangians.

The dimension 5 operators induce couplings to all the SM fields, 
\beq\label{eq:dim5}
\begin{split}
	\cL_{\rm int}^{(5)}
=&	\lambda_g \frac{\alpha_s}{4\pi m_S}S G^a_{\mu\nu}G^{a\mu\nu}+\lambda_B \frac{\alpha }{4 \pi c_W^2 m_S }S B_{\mu\nu}B^{\mu\nu} +
	\lambda_W \frac{\alpha}{4 \pi s_W^2 m_S}S W^a_{\mu\nu}W^{a\mu\nu}+\\
& -\frac{\lambda_H}{m_S} S (H^\dagger H)^2+\frac{\lambda_H'}{m_S} S D_\mu H^\dagger D^\mu H\\
&- \frac{\lambda_d}{m_S} S \bar Q_L d_R H- \frac{\lambda_u}{m_S} S \bar Q_L u_R H^c -\frac{\lambda_\ell}{m_S} S \bar L_L \ell_R H+ {\rm h.c.} \,, \\
\end{split}
\eeq
where $s_W\,(c_W)$ is the sine\,(cosine) of the weak mixing angle, $G^a_{\mu \nu}$, $B_{\mu \nu}$,  $W^a_{\mu \nu}$, are the QCD, hypercharge and weak isospin field strengths, respectively, $Q_L$ and $L_L$ are the quark and lepton left-handed doublets, respectively, and $d_R$, $u_R$, $\ell_R$ are the right-handed fields for down-type quarks, up-type quarks, and leptons. 
In general, the coefficients $\lambda_{d,u,\ell}$ are $3\times3$ complex matrices, where we do not display the dependence on the generational indices. However, since we are considering the CP conserving case, $\lambda_{d,u,\ell}$ are assumed real. 
The normalization of the operators in the first line of \eqref{eq:dim5} reflects the expectation that they are induced at 1-loop level. The $c_W=\cos\theta_W$ and $s_W=\sin\theta_W$ terms in the denominators ensure that the parameters  $\lambda_B$ and $\lambda_W$ do not contain the loop factors $g'^2/16\pi^2=\alpha/4\pi c_W^2$ and  $g^2/16\pi^2=\alpha/4\pi s_W^2$.  For simplicity we take $m_S$ as the operator normalization scale.  \footnote{Note that this does not mean that the EFT expansion is in $E/m_S$, but rather in $E/\Lambda$ with $\Lambda$ the scale of new states that is parametrically larger then $m_S$.}

After electroweak symmetry breaking~(EWSB), the $\mu_S SH^\dagger H$ term in \eqref{eq:dim4Lagr} and the $\lambda_H S (H^\dagger H)^2/m_S$ term in \eqref{eq:dim5} lead to the mixing between $S$ and $h$. In terms of the mass eigenstates, $h', S'$, we have $h=c_\alpha h'+s_\alpha S'$, $S=-s_\alpha h'+c_\alpha S'$, where the mixing angle is
\beq\label{eq:salpha}
s_\alpha\equiv \sin \alpha\simeq \frac{v^2}{m_S^2}\Big( \frac{\mu_S}{v}+\frac{v}{m_S}\lambda_H \Big) \,.
\eeq  
The searches for heavy Higgses decaying to $WW$ exclude $s_\alpha\lesssim 0.1$ \cite{Falkowski:2015swt}.\footnote{The bound on $\sin \alpha$ from WW resonance searches does not necessarily apply in this case, however, since the scalars can decay to SM fermions un-suppressed, reducing the branching fraction.}
In the following, we work to leading non-trivial order in the small parameter $v / m_S$. 

After EWSB the couplings of $S$ with the vector bosons are given by\footnote{From now on we drop the prime on the notation for the mass eigenstate (i.e., $S'\to S$ and $h' \to h$).} 
\beq
\begin{split}
	\cL_{\rm int}\supset
& 	\lambda_\gamma \frac{\alpha}{4 \pi m_S} S F_{\mu\nu} F^{\mu\nu}
	+\lambda_Z \frac{\alpha}{4 \pi m_S} S Z_{\mu\nu} Z^{\mu\nu}
	+\lambda_{Z\gamma} \frac{\alpha}{4\pi m_S} S Z_{\mu\nu} F^{\mu\nu}
	  \\
&	+\lambda_W \frac{\alpha}{2 \pi s^2_W m_S} S W^+_{\mu\nu} W^{-\mu\nu} +\frac{\kappa}{2} \Big(2\frac{m_W^2}{v}S W^{+\mu}W^-_\mu +\frac{m^2_Z}{v} S Z^\mu Z_\mu\Big) \, ,
\end{split}
\label{Lambda_int}
\eeq
where we have defined
\beq
	\kappa\equiv 2s_\alpha+ \lambda_H' v/m_S \, .
\eeq
Note that this parameter controls the decay rate to the longitudinal $W$ and $Z$ bosons. The coefficients in the first line in~\eqref{Lambda_int} may be written in terms of $\lambda_B$ and $\lambda_W$:
\begin{align}
	\label{eq:lambdagamma}
	\lambda_\gamma&=\lambda_B+\lambda_W \, , \\
	\label{eq:lambdaZgamma}
	\lambda_{Z\gamma}&=2\Big( \lambda_W\frac{c_W}{s_W}-\lambda_B\frac{s_W}{c_W}\Big) \, , \\
	\label{eq:lambdaZ}
	\lambda_Z&=\lambda_W\frac{c_W^2}{s_W^2}+\lambda_B\frac{s_W^2}{c_W^2}  \, .
\end{align}

The coupling of $S$ to a pair of Higgses is, to leading order in $v / m_S$, 
\begin{align}
{\cal L}_{\rm int}\supset -\Big(s_\alpha \frac{m_S^2}{v^2} + 2\lambda_H\frac{v}{m_S}  \Big)\frac{v}{2} S h^2 +\frac{\kappa-2s_\alpha}{2v} S \partial_\mu h \partial^\mu h\, , 
\end{align}
where $m_h=125$ GeV is the Higgs mass. These couplings mediate  $S\to hh$ decays, thus, non-observation of this decay channel may then be used to put constraints on $\kappa$, $s_\alpha$, and $\lambda_H$.

The couplings of $S$ to the SM fermions arise from the mixing with the Higgs and from dimension-5 operators in Eq.~\eqref{eq:dim5}. The couplings to up quarks are thus given by
\beq
{\cal L}_{\rm int}\supset -\Big(s_\alpha m_{u_i} \delta_{ij}+(\lambda_u)_{ij}\frac{v}{\sqrt2 m_S}\Big) S \bar u_{L,i} u_{R,j}+ {\rm h.c.},
\eeq
and similarly for the down quarks and charged leptons. The first term in the parenthesis is due to the mixing with the SM Higgs and is flavor diagonal. The couplings of dimension-5 operators can in principle be flavor violating, though such terms are tightly constrained \cite{Goertz:2015nkp}. 

In the case of pseudoscalar $S$ there is a smaller number of dimension-5 operators that we may write:
\beq\label{eq:dim5PS}
\begin{split}
	\cL_{\rm int}^{(5), PS}
=&	\tilde{\lambda}_g \frac{\alpha_s}{4\pi m_S}S G^a_{\mu\nu}\tilde{G}^{a\mu\nu}
	+\tilde{\lambda}_B \frac{\alpha }{4 \pi c_W^2 m_S }S B_{\mu\nu}\tilde{B}^{\mu\nu} 
	+\tilde{\lambda}_W \frac{\alpha}{4 \pi s_W^2 m_S}S W^a_{\mu\nu}\tilde{W}^{a\mu\nu}+\\
&- i\frac{\tilde{\lambda}_d}{m_S} S \bar Q_L d_R H- i\frac{\tilde{\lambda}_u}{m_S} S \bar Q_L u_R H^c - i\frac{\tilde{\lambda}_\ell}{m_S} S \bar L_L  \ell_R H+ {\rm h.c.}. 
\end{split}
\eeq
As for the scalar case, we assume CP conservation so that the $\tilde{\lambda}_{d,u,\ell}$ are real.

\subsection{Spin-0, $SU(2)$ doublet}

Another interesting possibility is that $S$ is one of the neutral components of a $SU(2)_L$ doublet with $Y= 1/2$ hypercharge. The scalar can in general mix with the SM Higgs doublet and the setup is captured by a general two Higgs doublet model. First, one is free to rotate the two Higgs doublets ($\Phi_{1,2}$) into a basis where only one obtains a vev
\beq
\Phi_1 = \left ( \begin{array}{c} G^+ \\ \frac{1}{\sqrt 2} (v+h_1 + i G^0) \end{array}
 \right ) \,, \quad \Phi_2 = \left ( \begin{array}{c} H^+ \\ \frac{1}{\sqrt 2} (h_2 + i h_3) \end{array}
 \right )\,.
 \eeq
The renormalizable scalar potential of the theory has the form
\beq
\mathcal L^{(4)}_{\rm scalar} = - \sum_{i,j=1,2} {\mu_{ij}^2}{} \Phi_i^\dagger  \Phi_j - \sum_{i,j,k,l=1,2} \lambda_{ijkl} \Phi_i^\dagger  \Phi_j \Phi_k^\dagger  \Phi_l + \rm h.c.\,,
\eeq
subject to the condition $\langle \Phi_2 \rangle =0$\,. In the CP conserving limit, i.e., assuming all parameters in the scalar potential to be real, the CP-odd pseudoscalar $h_3$ does not mix with the other neutral states and thus forms a mass eigenstate $h_3 \equiv A^0$. The two CP even scalars $h_{1,2}$,  on the other hand, do mix to form the mass eigenstates $h$ and $H^0$
\beq
 \left ( \begin{array}{c} h \\ H^0 \end{array} \right ) =  \left ( \begin{array}{cc} \cos \alpha & \sin\alpha\\ -\sin\alpha & \cos\alpha \end{array} \right )  \left ( \begin{array}{c} h_1 \\ h_2 \end{array} \right )\,.
\eeq
Close to the decoupling limit $\sin \alpha \ll 1$, one can identify $h$ with the observed Higgs boson at $m_h=125$~GeV and $H^0$ and/or $A^0$ with $S$. 
In this limit,
\beq
 m_{A^0} \simeq m_{H^0} \simeq m_{H^+} \simeq m_S,
 \eeq
 up to corrections of order $v^2/m_S^2 \simeq 0.1$\,.  The renormalizable interactions in the scalar potential lead to $H^0$ coupling to pairs of Higgs bosons ($A^0 h h$ coupling is forbidden by CP invariance). To leading order in $\sin \alpha$ it can be written as
\beq\label{eq:dim4Lagr2}
{\cal L}_{\rm scalar}^{(4)} \ni  -  v \left( {\lambda_{Hhh}^{(1)}}  + \lambda_{Hhh}^{(2)} \sin \alpha \right)H^0  h^2   \,,
\eeq
where $\lambda_{Hhh}^{(1,2)}$ are in general linear combinations of the $\lambda_{ijkl}$.  It is very important that this coupling of $H^0$ to two Higgses is tuned to be small in order not to violate constraints on the branching ratio of the resonance to $hh$. 

The renormalizable couplings of the scalars to SM fermions can be described as
 \beq
 \mathcal L^{(4)}_{f} = - \lambda^{}_{\Psi} \bar \Psi_L \Phi_1 \Psi_R - \lambda'_{\Psi} \bar \Psi_L \Phi_2 \Psi_R + \rm h.c.\,,
 \eeq
 where $\Psi_L = Q_L, L_L$ denotes the SM quark and lepton left-handed $SU(2)$ doublets, while $\Psi_R=u_R,d_R,\ell_R$ stands for the corresponding right-handed $SU(2)$ singlet fields of up-, down-quarks and charged leptons. The $\lambda^{(\prime)}_{d,u,\ell}$ are in general $3\times 3$ complex matrices. In the mass basis of SM fermions after EWSB, $\lambda_f = \sqrt 2 {\rm diag} (m_f) / v$. The $A^0, H^0$ couplings to fermions are given by
\begin{align}
 \mathcal L_{f}^{(4)} &= - \bar f_L^i f_R^j \left[  H^0\left( -\frac{m_i}{v} \delta^{ij} \sin\alpha + \frac{\lambda'_{ij}}{\sqrt 2} \cos\alpha \right) + i A^0 \frac{\lambda'_{ij}}{\sqrt 2}  \right]  + \rm h.c.\,,
 \end{align}
where $f=u,d,\nu,\ell$ and $\lambda'_{ij}$ are components of $\lambda_\Psi'$ given in the SM fermion mass basis. 

At operator dimension five, $\Phi_{1,2}$ only couple to lepton doublets 
\beq
\mathcal L_f^{(5)} = \sum_{i,j=1,2} \frac{\lambda^\nu_{ij}}{m_S} L_L^{c\dagger} \Phi^{c*}_i \Phi^{c \dagger}_j L_L + \rm h.c.\,.
\eeq  
The $\lambda^\nu_{11}$ term contributes to Majorana neutrino masses $m_\nu =   (\lambda^\nu_{11}+\lambda^{\nu\dagger}_{11})   v^2 / 2 m_S$ leading to a severe constraint $|\lambda^\nu_{11}| \lesssim 10^{-11}$. The $\lambda_{12,21}$, on the other hand, lead to $H^0$ couplings to neutrinos
\beq
\mathcal L_f^{(5)} \ni (\lambda^\nu_{12} + \lambda^\nu_{21})\frac{v}{m_S} \nu^T \nu H^0 + \rm h.c.\,.
\eeq  

Finally, direct couplings of $A^0$ to pairs of transverse gauge bosons are induced at dimension six (in the $s_\alpha\ll1$ limit these can be the leading contributions also for $H^0$).  The field strengths may couple to four independent combinations of scalar bilinears,
\es{operators}{
{\cal O}_1 &= \Phi_1^\dagger \Phi_1 \,, \qquad {\cal O}_2 = \Phi_2^\dagger \Phi_2 \,,\qquad 
{\cal O}_{3} = \Phi_1^\dagger \Phi_2 + \Phi_2^\dagger \Phi_1 \,, \qquad  {\cal O}_{4} = - i (\Phi_1^\dagger \Phi_2 - \Phi_2^\dagger \Phi_1) \,,
}
along with four additional variants, ${\cal O}^a_i$, that have $SU(2)$ generators $\tau^a$ inserted in-between the $\Phi$ fields (e.g., ${\cal O}^a_1 = \Phi_1^\dagger \tau^a \Phi_1$).
Assuming CP conservation the dimension-six couplings of $\Phi_1$, $\Phi_2$ to gauge bosons are then given by
\beq
\begin{split}
\mathcal L_{\rm gauge}^{(6)} &= \sum_{i = 1,2,3} \frac{ {\cal O}_i }{4\pi m_S^2} \left(  \lambda_{i}^g \alpha_s G_{\mu\nu} G^{\mu\nu} +  \lambda_{i}^B \frac{\alpha}{c_W^2} B_{\mu\nu} B^{\mu\nu} +  \lambda_{i}^W \frac{\alpha}{s_W^2} W_{\mu\nu} W^{\mu\nu}  \right) 
\\
& +\frac{ {\cal O}_4 }{4\pi m_S^2} \left(  \tilde \lambda_{4}^g \alpha_s G_{\mu\nu} \tilde G^{\mu\nu} +  \tilde\lambda_{4}^B \frac{\alpha}{c_W^2} B_{\mu\nu} \tilde B^{\mu\nu} + \tilde \lambda_{4}^W \frac{\alpha}{s_W^2} W_{\mu\nu} \tilde W^{\mu\nu} 
 \right) 
 \\
 & + \sum_{i = 1,2,3} \frac{\alpha  {\cal O}_i^a}{ 4\pi s_W c_W m_S^2} \lambda^{B\prime}_{i} W^a_{\mu \nu} B^{\mu\nu} + \frac{\alpha {\cal O}_4^a}{ 4\pi s_W c_W m_S^2} \tilde \lambda^{B\prime}_{4} W^a_{\mu \nu} \tilde B^{\mu\nu} \,.
 \end{split}
 \label{eq:dim6doublet}
\eeq
Hermiticity ensures that the $\lambda_i$ coefficients are real.
The leading operators that may generate decays of $H^0$ ($A^0$) to electroweak gauge bosons are ${\cal O}_3$ and ${\cal O}_3^a$ (${\cal O}_4$ and ${\cal O}_4^a$).
In principle, either $H^0$ or $A^0$ may be $S$ \footnote{Because of their mass degeneracy it is also possible that they both contribute to the diphoton signal leading to apparently wide resonant feature.}.  Since the calculations are similar for both scenarios, we assume in the following that the $\sim$750 GeV resonance $S$ is the scalar $H^0$.   Then, the couplings of $S$ to transverse electroweak bosons takes the same form as in~\eqref{Lambda_int}, except that now 
\begin{align}
	\label{eq:lambdagamma:doublet}
	\lambda_\gamma&=\frac{v}{m_S}\Big(\lambda_W+\lambda_B-\frac{1}{2} \lambda_B'\Big) \, , \\
	\label{eq:lambdaZgamma:doublet}
	\lambda_{Z\gamma}&=2\frac{v}{m_S}\Big( \lambda_W\frac{c_W}{s_W}-\lambda_B\frac{s_W}{c_W}-\frac{1}{4}\frac{c_W^2-s_W^2}{s_W c_W}\lambda_B' \Big) \, , \\
	\label{eq:lambdaZ:doublet}
	\lambda_Z&=\frac{v}{m_S}\Big(\lambda_W\frac{c_W^2}{s_W^2}+\lambda_B\frac{s_W^2}{c_W^2}+\frac{1}{2}\lambda_B'\Big)  \, ,
\end{align}
where
\beq
\lambda_W=-s_\alpha \lambda_1^W+c_\alpha \lambda_3^W, \qquad \lambda_B=-s_\alpha \lambda_1^B+c_\alpha \lambda_3^B, \qquad \lambda_B'=-s_\alpha\lambda_1^{B'}+ c_\alpha \lambda_3^{B'}.
\eeq
The parameter controlling the decays to longitudinal $W$ and $Z$, $\kappa$, is  unrelated to the above parameters, mirroring the singlet discussion.
 Note that, in this case, the couplings of $S$ to electroweak gauge bosons have one additional parameter compared to the $SU(2)_L$ singlet scenario.  This means that given two of the above couplings, the other two may be determined.

\subsection{Spin-0, $SU(2)$ triplet}

In this sub-section we introduce the effective Lagrangian assuming that the 750\,GeV resonance is the charge-neutral component, $S$, of an $SU(2)$ triplet, $T_S$. For simplicity we set the hypercharge to $Y_T=0$, so that $S$ is accompanied by two charge-1 components, $T_S^\pm$. The leading $T_S=T_S^i\sigma^i$ interactions with the SM are of dimension three and four,
 \begin{align}
 	\label{eq:Triplet34}
	 \mathcal L_{T}^{(3)} & = \mu_T H^\dagger T_S H\,, \\
	 \mathcal L_{T}^{(4)} & = \lambda_{1} H^\dagger T_S^2 H + \lambda_{2} H^\dagger H \, \tr \big( T^2_S \big)\,.
 \end{align}
The dimensionful parameter, $\mu_T$, induces a VEV for $T_S$ and is tightly constrained by EWPTs, $\mu_T/v < 1.6\,\%$~\cite{Agashe:2014kda,Yagyu:2013kva}. Consequently, its contribution to the $T_S\to hh$ decay rate is negligible.  The  $\lambda_{1,2}$ terms lead, after EWSB, to a universal mass shift of all $T_S$ components. The charged $T_S^\pm$ state is thus almost degenerate in mass with $S$. The $S$--$T^{\pm}$ mass splitting comes from the small mixing of $S$ with the higgs, \eqref{eq:Triplet34}, and from dimension six operator $ (H^\dagger T_S H)^2$\,.

 The couplings of $T_S$ to SM gauge bosons and fermions start at dimension five,
\beq
\begin{split}
	\label{eq:Triplet5}
	\mathcal L_{T}^{(5)} & = \lambda_{WB} \frac{\alpha}{4\pi m_S s_W c_W} {\rm tr} \big( T_S W_{\mu\nu}\big) B^{\mu\nu} 
	\\
&- \frac{\lambda_{d}}{m_S} \bar Q_L d_R T_S H - \frac{\lambda_{u}}{m_S} \bar Q_L u_R T_S H^c - \frac{\lambda_{\ell}}{m_S} \bar L_L \ell_R T_S H \,.
\end{split}
\eeq
where $W_{\mu\nu}=W_{\mu\nu}^a\tau^a$.
Note that at dimension 5, $T_S$ does not couple to  gluons. 
 At dimension 7 we find
\beq
\begin{split}
\label{eq:Triplet7}
\mathcal L_{T}^{(7)} & = \frac{\lambda_{W,1} \alpha}{4\pi m_S^3 c_W^2} H^\dagger W_{\mu\nu} T_S W^{\mu\nu} H  + \frac{\lambda_{W,2}\alpha}{4\pi m_S^3 c_W^2} H^\dagger   W_{\mu\nu}H\, \tr\big( W^{\mu\nu}  T_S\big) 
\\
& + \frac{\lambda_{W,3} \alpha}{4\pi m_S^3 c_W^2} H^\dagger   W_{\mu\nu} \tau^a H\, \tr \big( W^{\mu\nu} \tau^a  T_S\big) + \frac{\lambda_{W,4} \alpha}{4\pi m_S^3 c_W^2} H^\dagger  T_S W_{\mu\nu} W^{\mu\nu} H  + {\rm h.c.} 
 \\
& + \frac{\lambda_{W B, 1} \alpha}{4\pi m_S^3 s_W c_W} H^\dagger  H\, \tr \big( W^{\mu\nu}  T_S\big) B_{\mu\nu} + \frac{i \lambda_{W B, 2} \alpha}{4\pi m_S^3 s_W c_W} H^\dagger \tau^a H\, \tr \big( W^{\mu\nu} [\tau^a, T_S]\big)  B_{\mu\nu} 
\\
& + \frac{\lambda_{B } \alpha}{4\pi m_S^3 s_W^2 } H^\dagger  T_S H B_{\mu\nu} B^{\mu\nu}  + \frac{\lambda_{G } \alpha_s}{4\pi m_S^3 } H^\dagger  T_S H G_{\mu\nu} G^{\mu\nu} \,.
\end{split}
\eeq
It can be easily verified that the number of independent parameters is sufficient to completely de-correlate $S$ (and $T^{\pm}$) decay rates to various EW gauge boson final states.

For the triplet with $Y_T=1$ there is one renormalizable operator, $H^\dagger H T_S^\dagger T_S$. First nonrenromalizable interaction occur at dimension 7  where there are three operators of the form $\tr(H^T T_S H) B_{\mu\nu} B^{\mu\nu}$, $\tr(H^T T_S H) \tr(W_{\mu\nu}W^{\mu\nu})$, $(T_SH)^i (W_{\mu\nu}H)^i B^{\mu\nu}$. The analysis is thus similar to the case of the $S$ being part of the electroweak doublet.

\subsection{Spin-2}

Next we consider the spin-2 case. At leading dimension five operator level the most general interactions of a massive spin-2 field $S_{\mu\nu}$ satisfying the mass-shell conditions (c.f.~\cite{Buchbinder:1999ar}) with the SM can be described in terms of the traceless components of the energy momentum tensor\footnote{
Since $S^{\mu \nu}$ is traceless, it couples only to the traceless components of the SM stress tensors in \eqref{eq:Lspin2}.  For each of the SM scalars, vectors, and fermions, there is only a single dimension $4$ traceless, symmetric operator that may couple to $S^{\mu \nu}$.  In ~\eqref{eq:Lspin2} we chose these tensors to be conserved. However, there are no extra terms in the dimension $5$ effective Lagrangian for $S^{\mu \nu}$, beyond that written in Eq. \eqref{eq:Lspin2}, even if we assume that $S^{\mu \nu}$ does not necessary couple to conserved stress tensors.  
} 
\begin{align}
	\label{eq:Lspin2}
	\cL_{\rm int}^{\rm spin-2}
&=	\frac{S^{\mu\nu}}{m_S}\left[
	\sum_{A=W,B,G} \kappa_A\left( A_{\mu\alpha}A_{\nu\beta}g^{\alpha\beta} - \frac{g_{\mu\nu}}{4}A_{\alpha\beta}A^{\alpha\beta} \right) \right.\nonumber\\
	&\left . +\sum_f\frac{\kappa_f}{2}  \bar{f} \left( \gamma_\mu D_\nu + \gamma_\nu D_\mu \right)f + \kappa_H ( 4 D_\mu H^\dagger D_\nu H - g_{\mu\nu}  D_\alpha H^\dagger D^\alpha H  )
	 \right]\,.
\end{align}
After EWSB the gauge part becomes
\begin{align}
	\cL_{\rm int}^{\rm spin-2}
=	\frac{S^{\mu\nu}}{m_S}\Big[
&	\kappa_\gamma\left( F_{\mu\alpha}F_{\nu\beta}g^{\alpha\beta} - \frac{g_{\mu\nu}}{4}F_{\alpha\beta}F^{\alpha\beta} \right)
	+\kappa_{Z\gamma}\left( F_{\mu\alpha}Z_{\nu\beta}g^{\alpha\beta} - \frac{g_{\mu\nu}}{4}F_{\alpha\beta}Z^{\alpha\beta} \right) \nonumber\\
&	+\kappa_{Z}\left( Z_{\mu\alpha}Z_{\nu\beta}g^{\alpha\beta} - \frac{g_{\mu\nu}}{4}Z_{\alpha\beta}Z^{\alpha\beta} \right)
	+\kappa_{W}\left( W_{\mu\alpha}W_{\nu\beta}g^{\alpha\beta} - \frac{g_{\mu\nu}}{4}W_{\alpha\beta}W^{\alpha\beta} \right) \nonumber\\
	& + {\kappa_H} \left(    \frac{m_W^2}{v} W^+_\mu W^-_\nu + \frac{m_W^2}{v} W^+_\nu W^-_\mu + \frac{m_Z^2}{v} Z_\mu Z_\nu \right) \Big] \, ,
\end{align}
with 
\begin{align}
	\kappa_\gamma &= c^2_W \kappa_B + s^2_W \kappa_W \, , \\
	\kappa_{Z\gamma} &= 2c_W s_W(\kappa_W-\kappa_B) \, , \\
	\kappa_Z & = s^2_W \kappa_B + c^2_W \kappa_W \, .
\end{align}
The resulting relations among the $S$ decay amplitudes and rates to EW gauge bosons are exactly the same as in the scalar case discussed in Sec.~\ref{sec:spin0singlet}.  Note that $\kappa_{Z\gamma}$ vanishes exactly in the universal coupling limit $\kappa_i=\kappa$, where these interactions respect the local spacetime gauge symmetry.  This is to be contrasted with the spin-0 case in Eq.~\eqref{eq:lambdaZgamma}, where the vanishing of $\kappa_{Z\gamma}$ in general is not protected by a symmetry and thus requires fine-tuning of the $\kappa_B$ and $\kappa_W$ coefficients.

Finally, the generalization of the above results to higher $SU(2)_L$ representations of $S$ proceeds analogously to the spin-0 case (modulo different Lorentz contractions), although these are arguably less motivated from the theory perspective in the spin-2 case.

\section{Correlations between di-boson final states} \label{sec:correlations}

The observation of the 750\,GeV resonance, $S$, decaying to two photons generically implies that it should also decay to other pairs of electroweak bosons, $S\to WW, ZZ, Z\gamma$. The branching ratios for these decays are correlated, if the EFT expansion can be truncated at leading order. Some of these correlations have been discussed by authors focusing on loop-induced dimension-5 operators or in the  context of a specific model~\cite{Low:2015qho, Craig:2015lra}. Our only assumption is that the EFT described in section~\ref{sec:EFT} is valid and then phrase the correlations directly in terms of observables. 

We start with the case where $S$ is an electroweak singlet that does not couple to the Higgs, $\mu_S=\lambda_{S,H}=0$. In this limit there is no $S$--$h$ mixing and the mixing angle  \eqref{eq:salpha} vanishes, $s_\alpha=0$. It is then easy to rewrite the relations~\eqref{eq:lambdagamma}--\eqref{eq:lambdaZ} purely in terms of observables---the moduli of the decay amplitudes. We define the normalized decay amplitudes as
\beq\label{eq:Af}
\mathcal A_{f} \equiv \pm 4 \sqrt{\frac{\pi \Gamma_f}{m_S F_f} }\,,
\eeq
where $F_f$ are the small phase space correction factors due to massive final state particles, 
$F_{\gamma \gamma}:F_{\gamma Z} : F_{ZZ} : F_{WW} = 1:0.99 : 0.89 : 0.91$.

We work in the EFT limit where all the NP states that generate the dimension 5  operators in Eq.~\eqref{eq:dim5} are off-shell when running in the loop. The contributions from SM fermions are small. The only potentially significant contribution is from the top running in the loop, but even this is always sub-leading. The ratio of $S\to t\bar t$ decay width, $\Gamma_{t \bar t}$, and the $S\to \gamma\gamma$ rate induced entirely due to top quarks running in the loop, $(\Gamma_{\gamma\gamma})_{t\bar t}$, is $ (\Gamma_{\gamma\gamma})_{t\bar t}/\Gamma_{t \bar t}\simeq 5 \times 10^{-6}$~\cite{Franceschini:2015kwy, Angelescu:2015uiz, Aloni:2015mxa}. Using the bound from direct searches for $S\to t\bar t$ at 8\,TeV, $\Gamma(S\to {t \bar t})/\Gamma(S\to \gamma\gamma)\lesssim 300$, shows that the contribution to the $S\to \gamma\gamma$ rate from the top loop is always below ${\mathcal O}(10^{-3})$ and thus negligible.  
Due to chiral flip suppression, the $b$ and light quark contributions to $S\to \gamma\gamma$ are also always negligibly small, even if $S\to b\bar b$ or $S\to 2j$ were to saturate the total decay width, $\Gamma_S$.
In the limit where we can neglect top-quark contributions,  the $S\to \gamma\gamma$ decay amplitude and the $S\to ZZ, Z\gamma, WW$ decay amplitudes are all real. Namely, the coefficients $\lambda_{B,W}$ are real, while there are also no CP even (``strong'') phases from intermediate on-shell states. The decay amplitudes are thus determined up to  discrete choices of either a positive or a negative sign, as denoted in Eq.~\eqref{eq:Af}.

One can now identify four amplitude sum rules (any two of which are linearly independent) relating $\mathcal A_{\gamma\gamma}$ to $\mathcal A_{\gamma Z}$, $\mathcal A_{ZZ}$ and $\mathcal A_{WW}$, independent of the underlying parameters ($\lambda_{B,W}$) of the theory
\begin{align}
\mathcal A_{ZZ} c_W^2 + \mathcal A_{\gamma\gamma} s_W^2 &= \frac{c_W^2-s_W^2}{\sqrt 2} \mathcal A_{WW}\,, \label{sumrule1}
\\
\mathcal A_{WW} +\sqrt{2} \mathcal A_{\gamma\gamma}  &= \frac{c_W}{s_W} \mathcal A_{\gamma Z}\,, \\
\mathcal A_{ZZ} + \mathcal A_{\gamma\gamma}  &= \frac{c^2_W-s^2_W}{\sqrt 2 s_W c_W} \mathcal A_{\gamma Z} \,, \\
\mathcal A_{WW} - {}{\sqrt{2}} \mathcal A_{ZZ}  &= \frac{s_W}{c_W} \mathcal A_{\gamma Z} \,. \label{sumrule4}
\end{align}
It is clear that a non-vanishing $\mathcal A_{\gamma\gamma}$ implies that decays to at least two other final states should occur. If one additional $\mathcal A_f$ is measured or is tightly constrained, the remaining two rates are accurately predicted. Eqs.~\eqref{sumrule1}--\eqref{sumrule4} are valid in the limit of vanishing $\kappa$ and are correct for the transverse modes of the decay amplitude. For $\kappa\ne0$, the decay branching ratios to $ZZ$ and $WW$ will get contribution from the decay amplitudes to the longitudinal modes of the $W$ and the $Z$. Both of these are controlled by one additional parameter, $\kappa$.

Next, we move our discussion to the level of branching ratios, allowing $\kappa\ne0$.
The rates for $S$ decaying to two EW gauge fields are given by
\begin{align}\label{eq:Gammagammagamma}
	\Gamma_{\gamma\gamma}
=&	\lambda_\gamma^2 \frac{\alpha^2 m_S}{64\pi^3} \, , \\
	\Gamma_{Z\gamma}
=&	\lambda_{Z\gamma}^2 \frac{\alpha^2 m_S}{128\pi^3} \, , \\
	\Gamma_{ZZ}
=&	\lambda_Z^2 \frac{\alpha^2 m_S}{64\pi^3} 
	- 3\lambda_Z \kappa \frac{ \alpha  m^2_Z}{128\pi^2 v } 
	+\kappa^2  \frac{m_S^3 }{128\pi v^2} 
	+{\mathcal O}(m_Z^2/m_S^2)\, , \\
	\Gamma_{WW}
=&	\lambda_W^2 \frac{\alpha^2 m_S}{32\pi^3s^4_W} 
	- 3\lambda_W\kappa \frac{ \alpha  m^2_W}{64\pi^2v s^2_W}
	+\kappa^2 \frac{m_S^3 }{64\pi v^2}
	+{\mathcal O}(m_W^2/m_S^2)\, . \label{eq:GammaWW}
\end{align}
Above, we have kept the $m_{Z,W}^2/m_S^2$ suppressed terms explicit only in the interference term, proportional to $\kappa$. The rate for $hh$ is
\es{eq:Gammahh}{
\Gamma_{hh}  = {m_S \over 32 \pi} \sqrt{1 - 4{ m_h^2 \over m_S^2}} \left[ {m_S \over v} \left( {\kappa - 2 s_\alpha \over 2} \right) \left( 1 - 2{ m_h^2 \over m_s^2} \right) - {v \over m_S} \left( s_\alpha {m_S^2 \over v^2} + 2 \lambda_H {v \over m_S} \right) \right]^2 \,. 
}

The ratio $\Rrd_{Z\gamma}=\BR_{Z\gamma}/\BR_{\gamma\gamma}$ is controlled by a single parameter, 
\begin{align}
	r_{BW} \equiv \frac{\lambda_B}{\lambda_W} \, , 
\end{align}
while the ratios $\Rrd_{Z Z}=\BR_{ZZ}/\BR_{\gamma\gamma}$ and $\Rrd_{W W}=\BR_{WW}/\BR_{\gamma\gamma}$ are also sensitive to 
\begin{align}
	r_{\kappa W} \equiv \frac{\kappa}{\lambda_W} =  \frac{2s_\alpha+ \lambda_H' v/m_S}{\lambda_W} \, .
\end{align}
Note that $\kappa$ (and thus $r_{\kappa W}$) can be nonzero even if there is no mixing between $S$ and the Higgs. 
Measuring two out of three ratios of branching ratios fixes the two parameters, $r_{BW}$ and $r_{\kappa W}$. For instance, 
a measurement of $\Rrd_{Z\gamma}$ fixes $r_{BW}$, up to a two-fold ambiguity. A measurement of  either $\Rrd_{Z Z}$ or  $\Rrd_{WW}$ then determines $r_{\kappa W}$, which means that the remaining ratio of branching ratios is fully predicted. 

\begin{figure}[!t]
\centering
\includegraphics[width=0.45\textwidth]{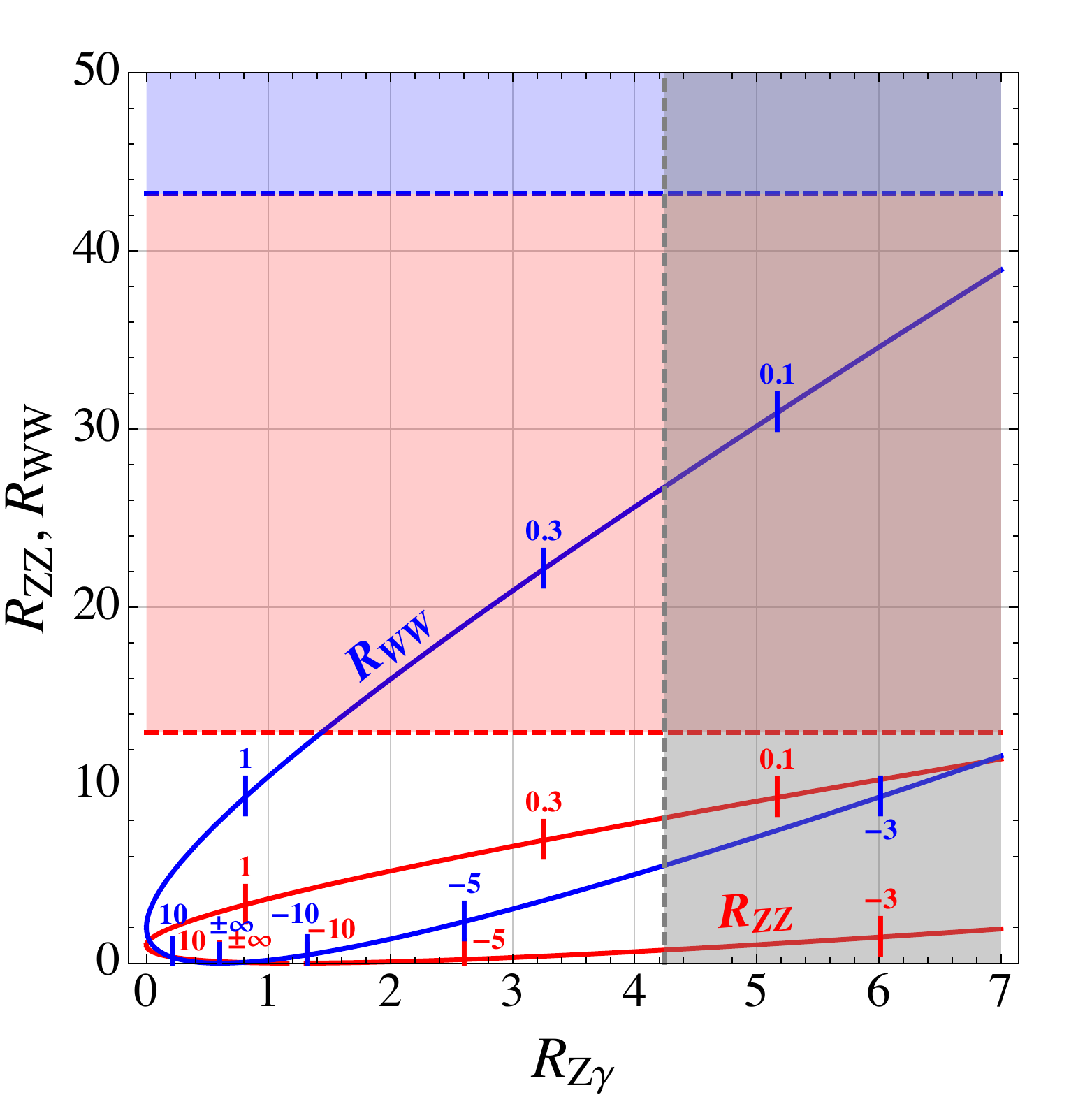}~~~
\includegraphics[width=0.48\textwidth]{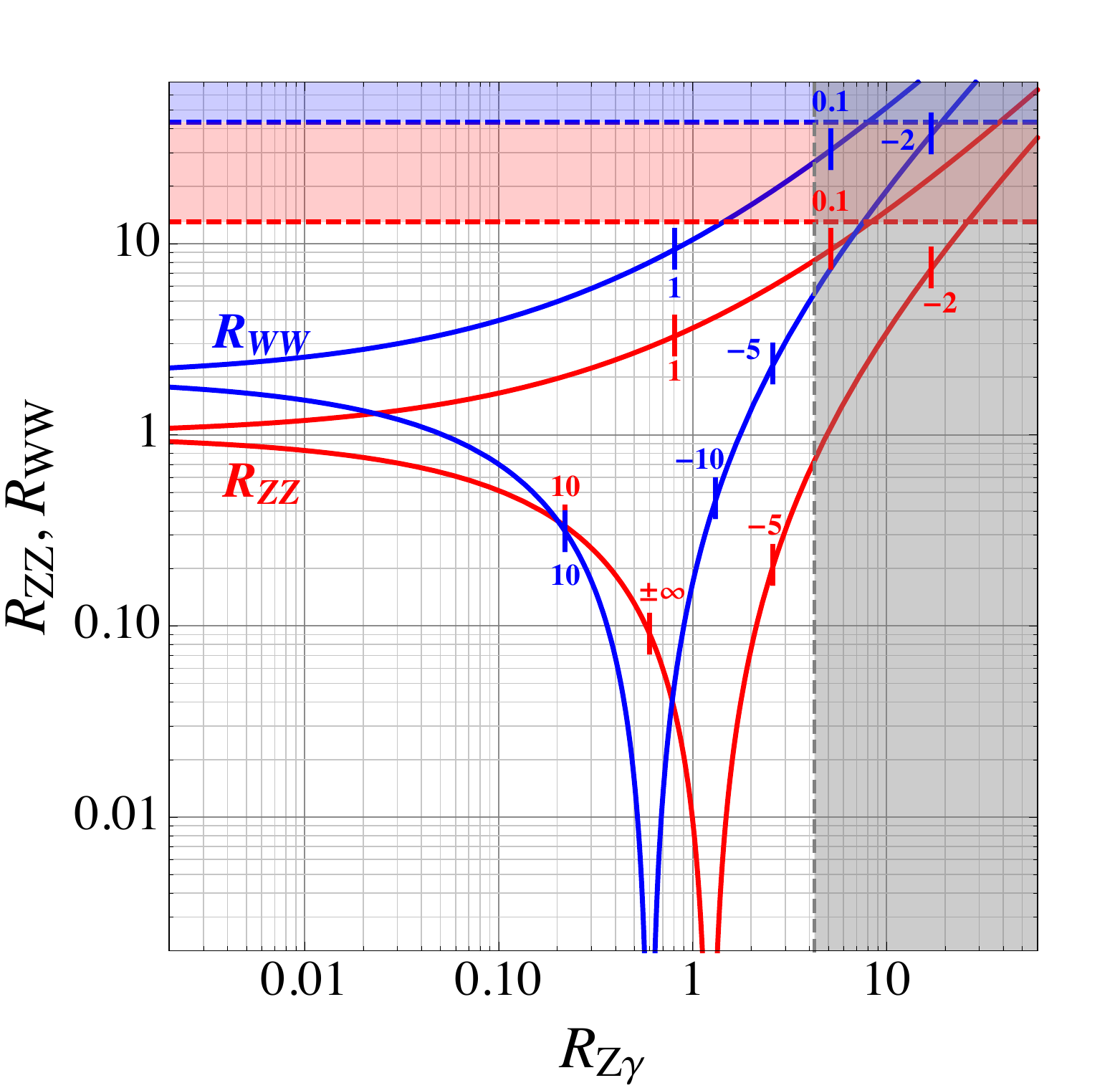}
\caption{$\Rrd_{ZZ}$~(red) and $\Rrd_{WW}$(blue) as function of $\Rrd_{Z\gamma}$ for $\kappa=0$. Left: linear scale, Right: log-log scale. The horizontal dashed lines are the current upper bounds on $\Rrd_{WW,ZZ}\lesssim 13, 43$, the vertical dashed line is  the upper bound $\Rrd_{Z\gamma}< 4.2$, appropriate for $b\bar b$ production \eqref{eq:RZabound}, while the ticks on the solid lines denote the corresponding values of $r_{BW}$.  For each $\Rrd_{Z \gamma}$ there are two solutions for $\Rrd_{ZZ, WW}$, shown by the lower and upper curves.  We denote the lower (upper) curves as branch 1 (branch 2).  }
\label{fig:correlationratioskappa0}
\end{figure}

\begin{figure}[!t]
\centering
\includegraphics[width=0.45\textwidth]{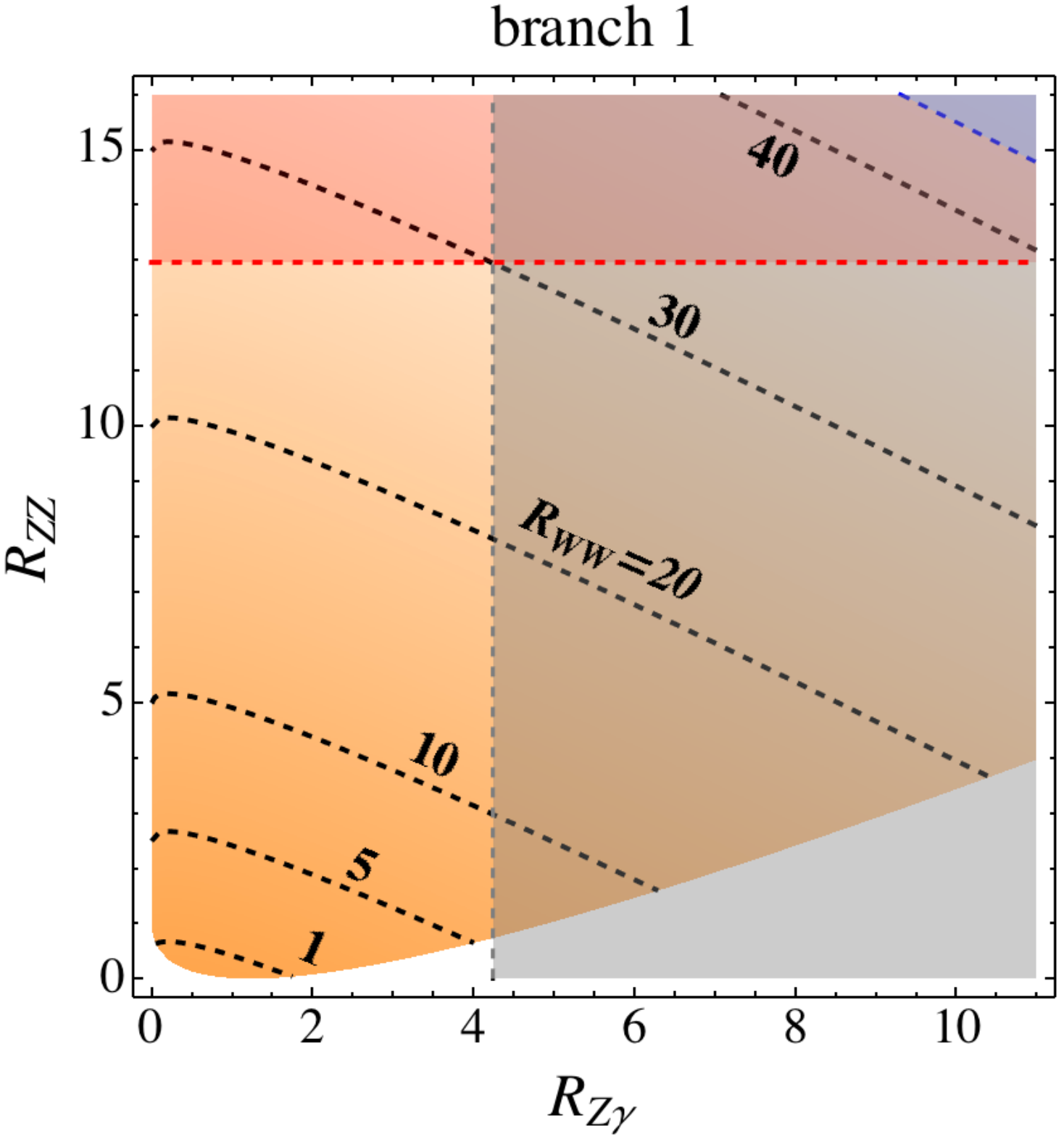}~~~
\includegraphics[width=0.45\textwidth]{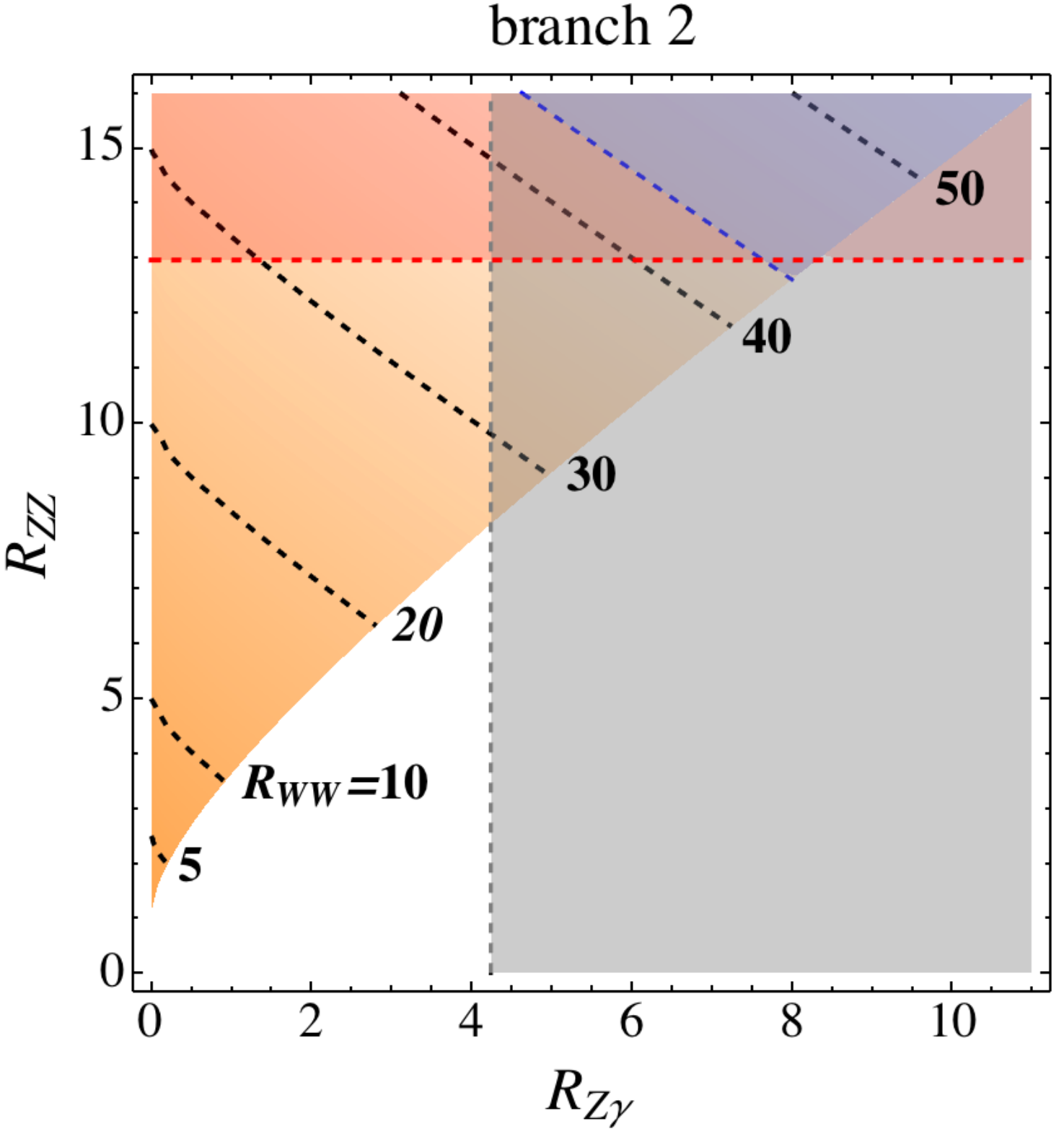}
\caption{Contours of $\Rrd_{WW}$ in the $\Rrd_{Z\gamma} - \Rrd_{ZZ}$ plane, with theoretically allowed region shaded orange. The two plots correspond to the two solutions, left\,(right) branch 1\,(2). The grey (red, blue) regions are excluded by $Z\gamma$ ($ZZ$, $WW$) resonance search. 
}
\label{fig:RWWyy}
\end{figure}

This is illustrated in Figures \ref{fig:correlationratioskappa0} and \ref{fig:RWWyy}.
In Fig. \ref{fig:correlationratioskappa0} we show the correlations for the $\kappa=0$ limit (see also Ref.~\cite{Low:2015qho}). In this case all three ratios, $\Rrd_{Z \gamma}$, $\Rrd_{Z Z}$, $\Rrd_{W W}$, depend on only one parameter, $r_{BW}$. This we trade for $\Rrd_{Z\gamma}$ and show $\Rrd_{Z Z}$, $\Rrd_{W W}$ as functions of $\Rrd_{Z\gamma}$. Current collider data imply  $\Rrd_{Z\gamma}\lesssim 4.2\,(6.6 \,{\rm fb}/\sigma_{\gamma\gamma})$, see Eq.~\eqref{eq:RZabound} above, from which we find 
\begin{align}
	\Rrd_{ZZ} \lesssim  19 - 6.3 \, , \qquad
	\Rrd_{WW} \lesssim 68 - 20  \, ,
\end{align}
assuming $\sigma_{\gamma\gamma} = 2- 10 $~fb. For large $\sigma_{\gamma\gamma}$ values,  these bounds are thus comparable to the 8\,TeV constraints from Eqs.~\eqref{eq:RWWbound}-\eqref{eq:Rhhbound}.

It is possible that one of the decays, $S\to ZZ, WW, Z\gamma$ has vanishing branching ratio. It is, however, impossible for two of them to vanish simultaneously. For instance, for vanishing $S\to Z\gamma$,  $\Rrd_{Z\gamma}=0$, one has $\Rrd_{ZZ}=1$ and $\Rrd_{WW}=2$. If $\Rrd_{ZZ}=0$, then $\Rrd_{Z\gamma}\approx 1.2$ and  $\Rrd_{WW}\approx 0.37$ or $\Rrd_{WW}\approx 12$. For $\Rrd_{WW}=0$, one has $\Rrd_{Z\gamma}\approx 0.6$ with either $\Rrd_{ZZ}\approx 0.09\,$ or $\Rrd_{ZZ}\approx 2.9\,$. We conclude that  in the limit of $\kappa=0$ at least two of the ratios should deviate from zero, in accordance with amplitude sum rules in \eqref{sumrule1}--\eqref{sumrule4}. For $\Rrd_{Z\gamma}\simeq 1$ the two remaining ratios, $\Rrd_{Z Z}$ and $\Rrd_{WW}$, are suppressed, see Fig.~\ref{fig:correlationratioskappa0} right.  Nevertheless, the  $\Rrd_{Z Z}$ and $\Rrd_{WW}$ are never zero simultaneously. Finally, note that establishing upper bounds 
\beq
\Rrd_{ZZ,WW}\lesssim0.4~~{\rm and}~~\Rrd_{Z\gamma}\lesssim0.2,
\eeq
 would exclude the $\kappa=0$ case.

Next, we discuss the general case keeping $\kappa\ne0$. In Fig.~\ref{fig:RWWyy} we show contours of $\Rrd_{WW}$ in the $\Rrd_{Z\gamma} - \Rrd_{ZZ}$ plane. There is a two-fold ambiguity when solving for $\Rrd_{WW}$ in terms of $\Rrd_{Z\gamma} - \Rrd_{ZZ}$. The two panels in Fig.~\ref{fig:RWWyy} show the two solutions for $\Rrd_{WW}$, which we call branch~1 and branch~2. We see that when  $\Rrd_{Z\gamma}=0$ both $\Rrd_{Z Z}$ and $\Rrd_{W W}$ need to be nonzero. Similar conclusions apply if $\Rrd_{Z Z}=0$ or $\Rrd_{WW}=0$. This means that even for $\kappa\ne 0$ 
at least two of the ratios, $\Rrd_{Z \gamma}$, $\Rrd_{Z Z}$, and $\Rrd_{WW}$, need to be nonzero. 

We re-emphasize that the above correlations between different di-boson final states rely on a valid EFT expansion. The assumption is that one can truncate the EFT at dimension 5 operators, while higher orders are neglible. It is possible, for instance, to have a positive signal only in the diphoton channel, if the interactions of $S$ with the EW gauge fields are mediated by a dimension 9 operator,
\beq
{\cal L}_9\supset\frac{1}{m_S^5} S (g B_{\mu\nu} H^\dagger H - g' W_{\mu\nu}^i H^\dagger \sigma^i H)^2 \,.
\eeq
It is not easy to see what symmetry would allow this dimension 9 operator, but forbid lower dimensional operators that we were discussing so far. 
If only a diphoton signal is observed with no indication of decays to other di-boson states, this would signal a breakdown of EFT. 

Note that our conclusions apply also to the case where $S$ is a pseudo-scalar. The effective interaction with the SM gauge field are given in Eq.~\eqref{eq:dim5PS}. The different ratios, $\Rrd_{Z\gamma,ZZ,WW}$ are all controlled by one ratio of parameters, $\tilde \lambda_B/\tilde \lambda_W$. The discussion is thus the same as in the case of a scalar $S$, but with $\kappa=0$. Therefore,  the correlations shown in Fig.~\ref{fig:correlationratioskappa0} also hold for the pseudoscalar case. 

At this point it is instructive to consider whether these conclusions derived for the case of a $SU(2)_L$ singlet $S$ can be invalidated if $S$ is part of a larger weak multiplet. If $S$ is one of the neutral states of a doublet ($H_S$), the leading interactions with SM gauge bosons arise purely from the $S-h$ mixing.  Additional interactions of $S$ with gauge bosons arise at dimension $6$, as shown in~\eqref{eq:dim6doublet}.  Through dimension $6$, there are five relevant parameters that go into determining the $3$ ratios $\Rrd_{WW}$, $\Rrd_{ZZ}$, and $\Rrd_{Z \gamma}$.  These are  $\lambda_W$, $\lambda_B$, $\lambda_B'$, $\sin \alpha$, and an analog of the $\lambda_H'$ singlet term, arising from a dimension $6$ operator $\sim$$(H D_\mu H)^2$.  However, it may be verified that---just as in the singlet case---$\sin \alpha$ and $\lambda_H'$ only enter into the ratios in the form of a single parameter $\kappa$.  When $\kappa = 0$, then it may be seen directly through~\eqref{eq:lambdagamma:doublet}-\eqref{eq:lambdaZ:doublet} that there is not enough freedom to set more than two of ratios to zero.  That is, in addition to $\Gamma_{\gamma \gamma}$, at least one of the other electroweak final states must be non-vanishing.  A direct analysis shows that this conclusion is not changed when $\kappa \neq 0$, in analogy to the singlet case.

More interesting is the case of a $SU(2)_L$ triplet ($T_S$) with hypercharge $Y_T = 0$ or $\pm 1$. Consider for concreteness $Y_T=0$, where its interaction with the SM (up to mass dimension~5) are given in Eqs.~\eqref{eq:Triplet34}--\eqref{eq:Triplet5}. As we can see the leading interaction between $T_S$ and EW gauge boson is controlled by only one parameter, $\lambda_{WB}$. This means that the branching ratios to $S\to ZZ, Z\gamma, WW$ are all uniquely predicted in terms of the $\BR_{\gamma\gamma}$. The corresponding couplings in \eqref{eq:Gammagammagamma}--\eqref{eq:GammaWW} are given by 
\beq
\lambda_{\gamma} : \lambda_{\gamma Z} : \lambda_Z : \lambda_{W} = s_W c_W : c_W^2-s_W^2 : - s_W c_W : 0, \qquad {\rm Triplet}, Y_T=0.
\eeq
 The predicted ratio $\Rrd_{Z\gamma} \simeq (c_W/s_W)^2 \simeq 3.3$ is in slight tension with the existing LHC searches for $Z\gamma$ resonances. The tight correlations among $\lambda_{f}$ are lifted at dimension seven,  see $\mathcal L_{T}^{(7)}$ in Eq.~\eqref{eq:Triplet7}, thus again necessarily implying violation of the EFT power counting. It can be easily verified that the number of independent parameters at that order is sufficient to completely de-correlate $S$ 
 decay rates to various EW gauge boson final states.
 
 If $S$ is part of an electroweak triplet this means that that $T^{\pm}$ charged states are also being produced. The production cross sections depend on the couplings of the triplet to the light quarks, $b$ quarks and to the electroweak bosons.   The dominant decays are $T^\pm\to 2j $ and the decays to gauge bosons, $T^{\pm} \to W^{\pm} \gamma$ and  $T^{\pm} \to W^{\pm} Z$.

\section{VBF Production} \label{sec:VBF}

\begin{figure}[!t]
\centering
\includegraphics[width=0.95\textwidth]{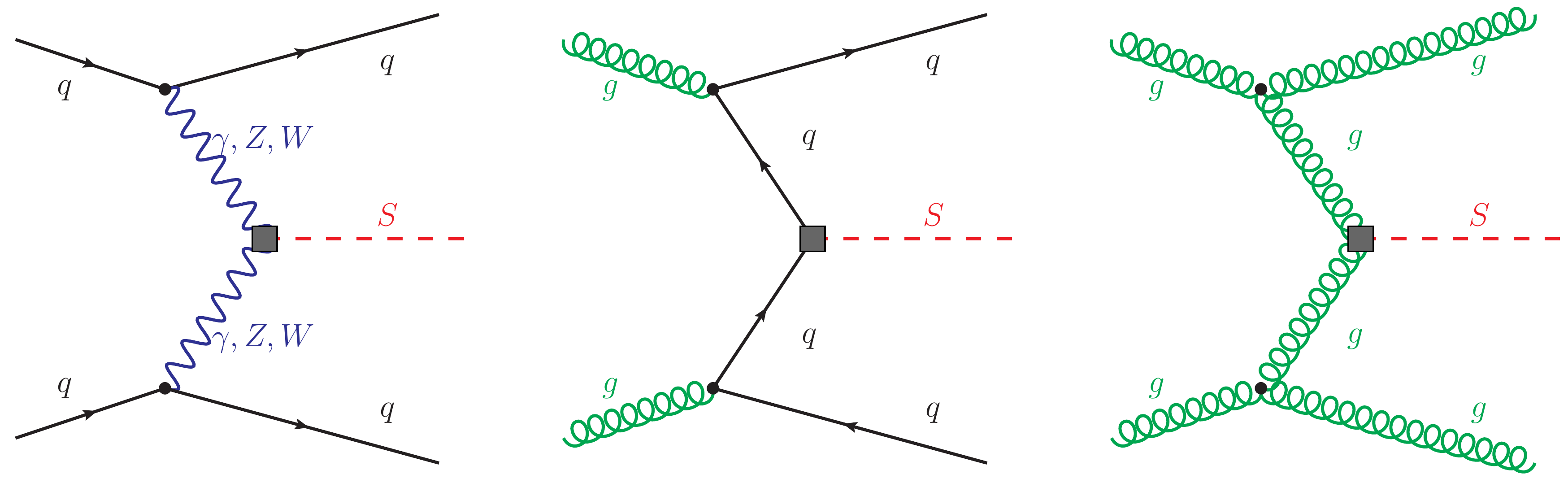}
\caption{Examples of leading order diagrams contributing to the VBF-tagged production of $S$ due to its couplings to EW gauge bosons (left-hand side), SM quarks (center) and gluons (right-hand side).}
\label{fig:vbf}
\end{figure}

An interesting possibility to probe different couplings of the resonance to the standard model is given by measuring the  production in association with two forward jets, similar to the Higgs vector boson fusion.\footnote{Note that the discussion presented here is related but complementary to that in~\cite{Csaki:2016raa}.  In that work, it was shown that a variety of related observables, such as the jet multiplicity and kinematic distributions, may be used to distinguish the gluon fusion and photon fusion scenarios.}

Given that the new state is probed in the $\gamma\gamma$ final state together with the unavoidable correlations between the other electroweak final states, we learn that there is a minimal contribution to the VBF cross section.  In practice, we may search for VBF by choosing a set of VBF cuts, such as requiring two jets with: 
\beq
p_T(j) > 60\,{\rm GeV}\,,\quad |\eta_j|<5\,,\quad |\Delta \eta (jj)| > 3.6\,, \quad m_{jj} > 1\,{\rm TeV}\,,
\eeq
while not imposing any cuts on the final state photons. Using MadGraph simulations for both spin hypotheses and all production modes we checked that imposing photon cuts in addition to the VBF ones leads to similar efficiency corrections as in the inclusive case.
Note that all production channels will generate events that pass the VBF cuts (see Fig.~\ref{fig:vbf}), though we expect a larger ratio of the rate after the VBF cuts versus the inclusive rate for actual vector-boson fusion production channels. For simplicity, we restrict this discussion to spin $0$, and we comment briefly on spin $2$ at the end of the section.  
  
Below, we use MadGraph~\cite{Alwall:2014hca} to simulate the different cross sections. For spin zero the cross section for events passing the VBF cuts is given by  
\begin{align}
	\label{eq:VBFXsec0}
	\sigma_{\rm VBF, 0}^{13\,{\rm TeV}} 
= 	  \big(  
	&1  + 1.1 \Rrd_{Z\gamma} +  0.76\Rrd_{ZZ} +1.2  \Rrd_{WW} 
	\pm0.46\sqrt{\Rrd_{ZZ}}\pm1.0\sqrt{\Rrd_{Z\gamma}}\pm1.0 \sqrt{ \Rrd_{Z\gamma} \Rrd_{ZZ} } \nonumber\\
	&+ 93 \Rrd_{gg}+2.3\Rrd_{u\bar{u}}+1.0\Rrd_{d\bar{d}}+0.11\Rrd_{s\bar{s}} \nonumber \\
	&+ 0.084\Rrd_{c\bar{c}}+0.058\Rrd_{b\bar{b}}   
	\big)\BR_{\gamma\gamma}    \left( \frac{  \Gamma_S}{\rm 45\, GeV} \right)   42 \, {\rm fb}  \, ,
\end{align}
where the $\pm$ stands for considering both interference possibilities. Here  $\Rrd_{p}$ are the signal rates in channel $p$ normalized to the diphoton rate, cf. Eq. \eqref{eq:Ri:definition}. In \eqref{eq:VBFXsec0} we quote only the central values for the numerical factors obtained using NLO NNPDF 2.3~\cite{Ball:2014uwa} pdf set. The errors are expected to be at the level of tens of percent, depending on the channel.
The inclusive cross section is given by~\cite{Franceschini:2015kwy} 
\begin{align}
	\label{eq:INCXsec0}	
\sigma_{\rm inc, 0}^{13\,{\rm TeV}} 
= 	\left( \frac{\Gamma_S}{\rm 45\, GeV} \right) \BR_{\gamma\gamma} \big(  
	& 64 \Rrd_{gg}+36\Rrd_{u\bar{u}}+21\Rrd_{d\bar{d}}+2.8\Rrd_{s\bar{s}}+1.2\Rrd_{c\bar{c}}+0.51\Rrd_{b\bar{b}}   \nonumber\\
	&+1  + 0.36 \Rrd_{Z\gamma} +  0.060\Rrd_{ZZ} +0.092  \Rrd_{WW}
	\big)  \, 4.1 \, {\rm pb} \, ,
	\end{align}
where the interference terms are negligible.  For the inclusive photon fusion cross section estimate we use  \cite{Harland-Lang:2016qjy} which was obtained using the MMHTNLO pdf set \cite{Harland-Lang:2014zoa} with NLO (in $\alpha_S$) DGLAP running. The  related error was estimated in \cite{Harland-Lang:2016qjy} to be $\sim \pm 15-20\%$. The inclusive rates for the remaining production channels were obtained using NLO NNPDF 2.3~\cite{Ball:2014uwa} pdf.
In the case of $SWW$, $SZ\gamma$ and $SZZ$ couplings, a significant part of the inclusive production is due to associated production of $pp\to S W$ and $pp\to S Z$. Searching for associated production is one way of probing the $S$ couplings to vector bosons.
We focus on the VBF part. In particular, we consider the ratio between the VBF and the inclusive production cross sections
\begin{align}
	R_{\rm VBF/inc} \equiv  \frac{\sigma_{\rm VBF}^{13\,{\rm TeV}}}{\sigma_{\rm inc}^{13\,{\rm TeV}}} \, .
\end{align}

\begin{figure}[!t]
\centering
\includegraphics[width=0.45\textwidth]{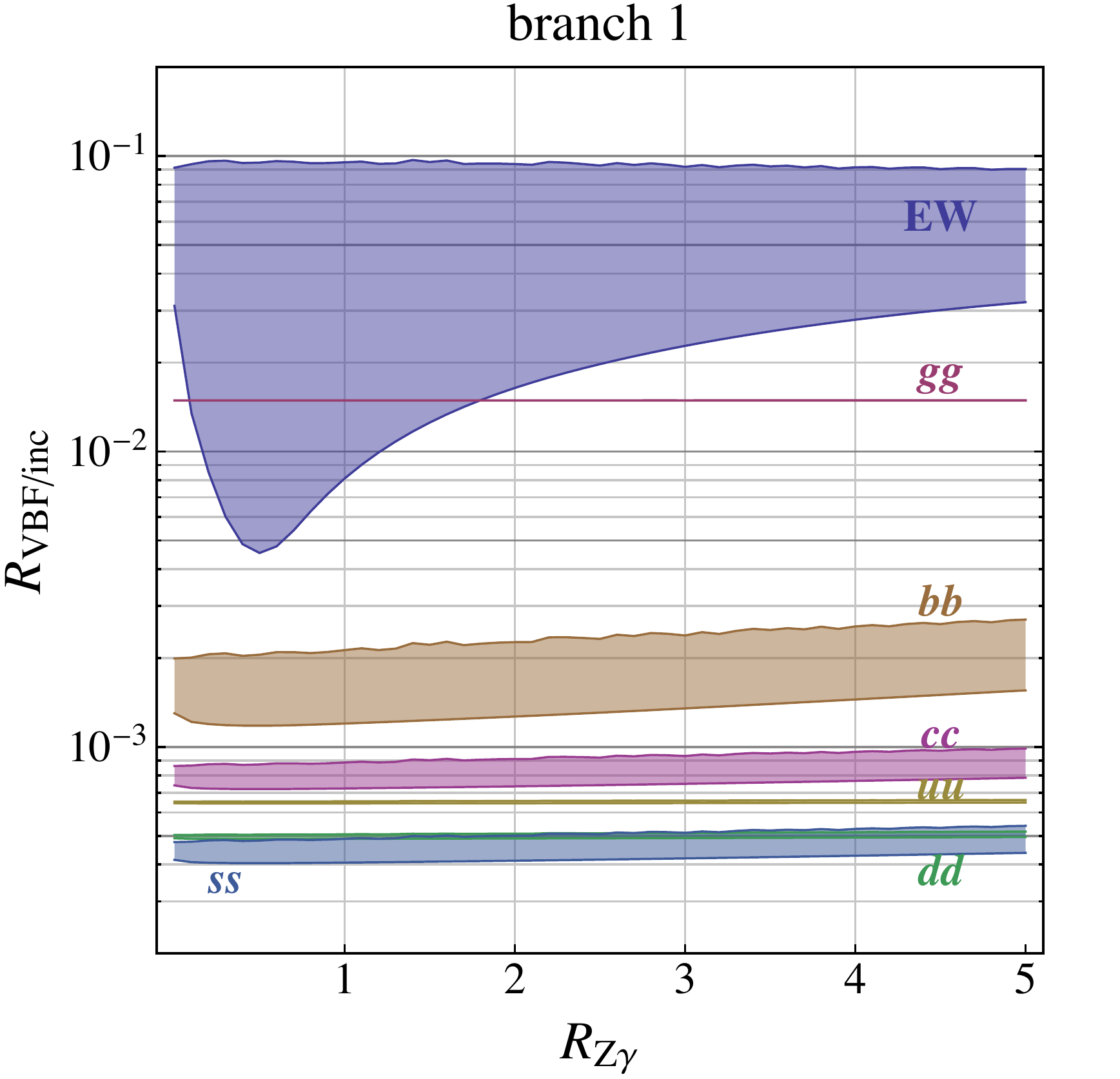}
\includegraphics[width=0.45\textwidth]{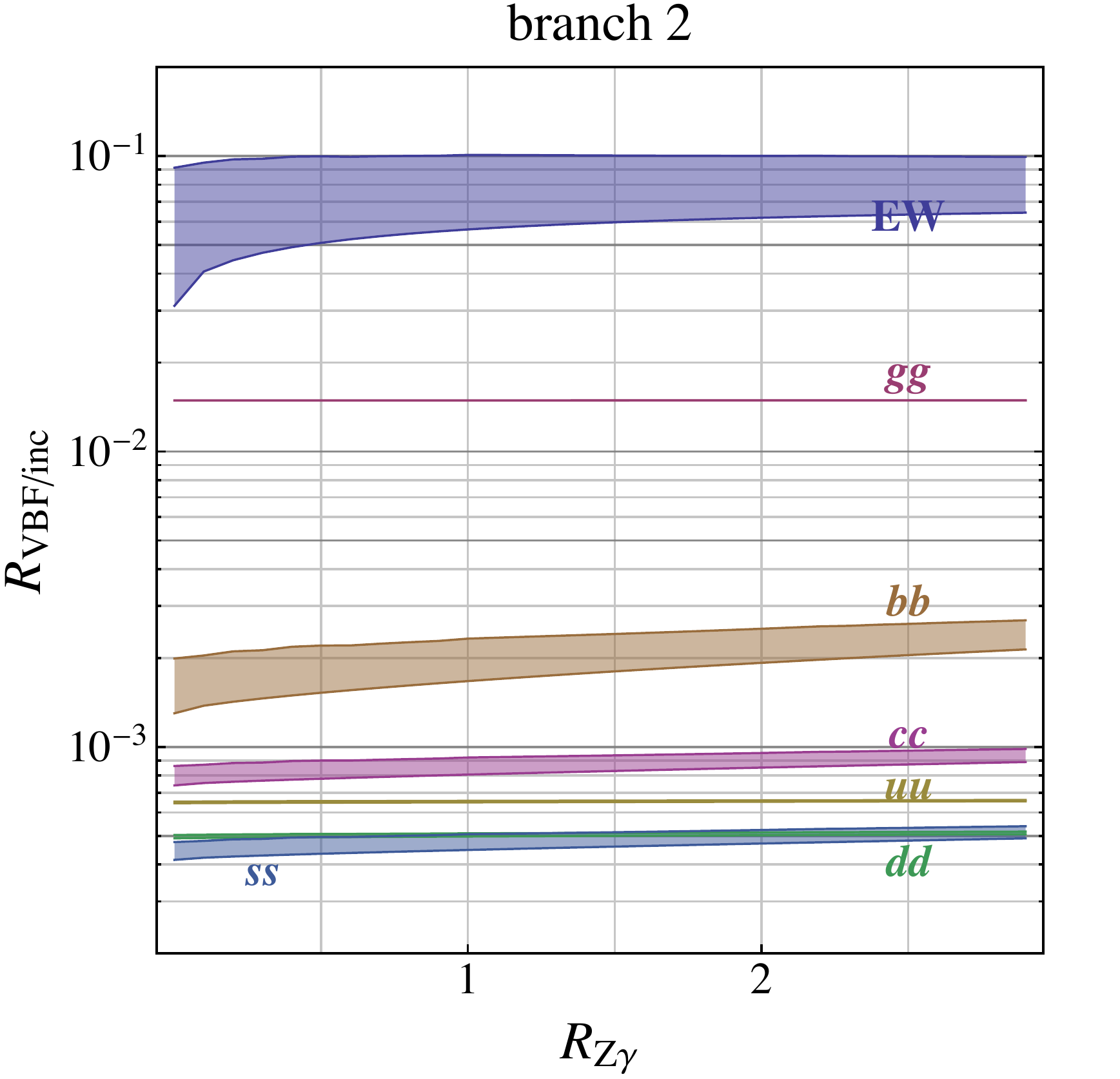}
\caption{$R^{0}_{\rm VBF/inc}$ for the different production mechanisms. Left: branch 1 for $\Rrd_{WW,ZZ}$, Right: branch 2 for $\Rrd_{WW,ZZ}$ (See Fig.~\ref{fig:correlationratioskappa0}). 
}
\label{fig:RVBFinc}
\end{figure}

 In Fig.~\ref{fig:RVBFinc} we plot $R_{\rm VBF/inc}$ as a function of $\Rrd_{Z \gamma}$ for a variety of quark and gluon production channels.  We switch on one production channel at the time,  $p=gg,u\bar{u},d\bar{d},s\bar{s},c\bar{c},b\bar{b}$, and take $\Rrd_p$ to be the maximal experimentally allowed value, given in~\cite{Franceschini:2015kwy}. In addition we allow for the electroweak production. For fixed $\Rrd_{Z\gamma}$ the diphoton rates \eqref{eq:VBFXsec0}, \eqref{eq:INCXsec0} still depend on two variables that can be taken as $\Rrd_{ZZ}$ and $\kappa$. These are varied making sure that the 8\,TeV bound on $\Rrd_{WW,ZZ}$ are obeyed, resulting in the colored regions in Fig.~\ref{fig:RVBFinc} (left panel for branch-1, right panel for branch-2).   
 
 The purely electroweak production is shown as a blue band. The VBF to inclusive ratio, $R_{\rm VBF/inc}$, is seen to depend heavily on whether $\Rrd_{WW}$ and $\Rrd_{ZZ}$  are from the upper or  the lower branch solution (for a given $\Rrd_{Z \gamma}$, see, for example,~Fig.~\ref{fig:correlationratioskappa0}).  
For the other production channels the width of the bands is smaller; the smaller the width of the bands, the smaller the relative contribution of the VBF production from electroweak gauge boson fusion. For instance, for the maximal value of gluon fusion, the VBF production is always sub-leading. For $b\bar b$ production, on the other hand, electroweak gauge boson production can be comparable. 

  Interestingly, we find that for $\Rrd_{Z \gamma}\gtrsim2.5$  the observable $R_{\rm VBF/inc}$ can distinguish the pure electroweak production from all the other production channels.  In particular, if $R_{\rm VBF/inc}$ is found to be greater than 2.6\%, then the production must occur through electroweak gauge boson fusion.  If $R_{\rm VBF/inc} <0.3\% $, then the production must receive contributions from the quark channels.  This is particularly relevant given that in Sec.~\ref{sec:fit} we showed that the current most-favored production channels are heavy-quark annihilation and gluon fusion; the VBF analysis gives a method for discriminating these scenarios.   The ratio $R_{\rm VBF/inc}$ cannot be greater than 5.3\% in any of the channels. 
  
The ability for $R_{\rm VBF/inc}$ to distinguish between channels is not as clear in the spin-$2$ scenario, for one because in this case the quark contributions to VBF are enhanced relative to the spin-$0$ scenario.  Explicitly, we find that 
\begin{align}
	\label{eq:VBFXsec2}
	\sigma_{\rm VBF, 2}^{13\,{\rm TeV}} 
= 	  \big(  
	&1  + 0.97 \Rrd_{Z\gamma} +  0.59\Rrd_{ZZ} +0.90  \Rrd_{WW} 
	\pm0.18\sqrt{\Rrd_{ZZ}}\pm0.95\sqrt{\Rrd_{Z\gamma}} \nonumber\\
	&\pm0.68 \sqrt{ \Rrd_{Z\gamma} \Rrd_{ZZ} }+ 66 \Rrd_{gg}+49\Rrd_{u\bar{u}}+19\Rrd_{d\bar{d}}+1.5\Rrd_{s\bar{s}} \nonumber \\
	&+ 0.95\Rrd_{c\bar{c}}+0.63\Rrd_{b\bar{b}}   
	\big)\BR_{\gamma\gamma}    \left( \frac{  \Gamma_S}{\rm 45\, GeV} \right)   350 \, {\rm fb}  \, ,
\end{align}
while 
\begin{align}
	\label{eq:INCXsec2}
	\sigma_{\rm inc, 2}^{13\,{\rm TeV}} 
= 	\left( \frac{\Gamma_S}{\rm 45\, GeV} \right) \BR_{\gamma\gamma} \big(  
	& 39 \Rrd_{gg}+19\Rrd_{u\bar{u}}+12\Rrd_{d\bar{d}}+1.3\Rrd_{s\bar{s}}+0.63\Rrd_{c\bar{c}}+0.27\Rrd_{b\bar{b}}   \nonumber\\
	&+1  + 0.53 \Rrd_{Z\gamma} +  0.16\Rrd_{ZZ} +0.21  \Rrd_{WW}
	\big)  \, 38 \, {\rm pb} \,.
\end{align}
The enhanced $R_{\rm VBF/inc}$ ratio for quark production for spin $2$ compared to spin $0$ may be understood because the coupling of $S_{\mu \nu}$ to quarks scales with the momentum of the quarks, while the coupling of the spin-$0$ resonance to quarks is momentum independent.  As such, the $S_{\mu \nu}$ production through quark anti-quark production occurs preferentially with higher quark momenta and thus a larger fraction of the events pass the VBF cuts, compared to spin-$0$ production.  Thus, for spin $2$ the VBF cuts are less efficient at reducing the contribution from quark anti-quark annihilation.

\section{Constraints on the partial widths of $S$} \label{sec:SMrate}

One of the more exciting possibilities is that the low-energy effective field theory of the SM plus a single new $\sim$750 GeV particle is insufficient at low energies to explain the observations and that new light states are also required.  For example, there are hints that the total width $\Gamma_S$ may be quite large, $\sim$45 GeV.  Such a large width may be obtained within the EFT framework described in this paper, but---as we show below---obtaining this width puts strong constraints on other observables, such as $R_{\rm VBF/inc}$, or may require large branching ratios to $t \bar t$.  However, it should be noted that the large width is only supported by ATLAS data~\cite{ATLAS-CONF-2015-081} at the moment.

\begin{figure}[!t]
\centering
\includegraphics[width=0.45\textwidth]{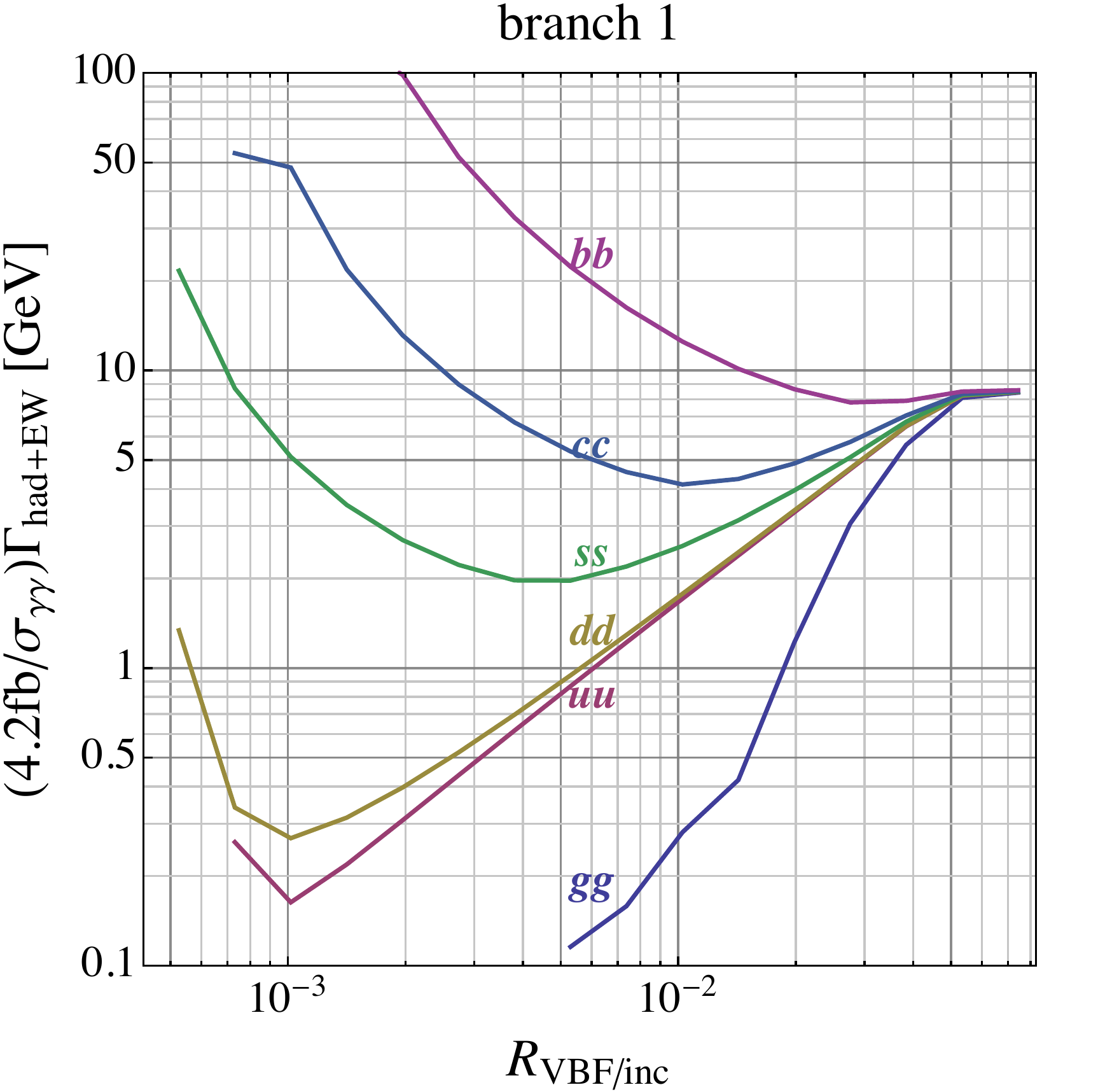}~~~
\includegraphics[width=0.45\textwidth]{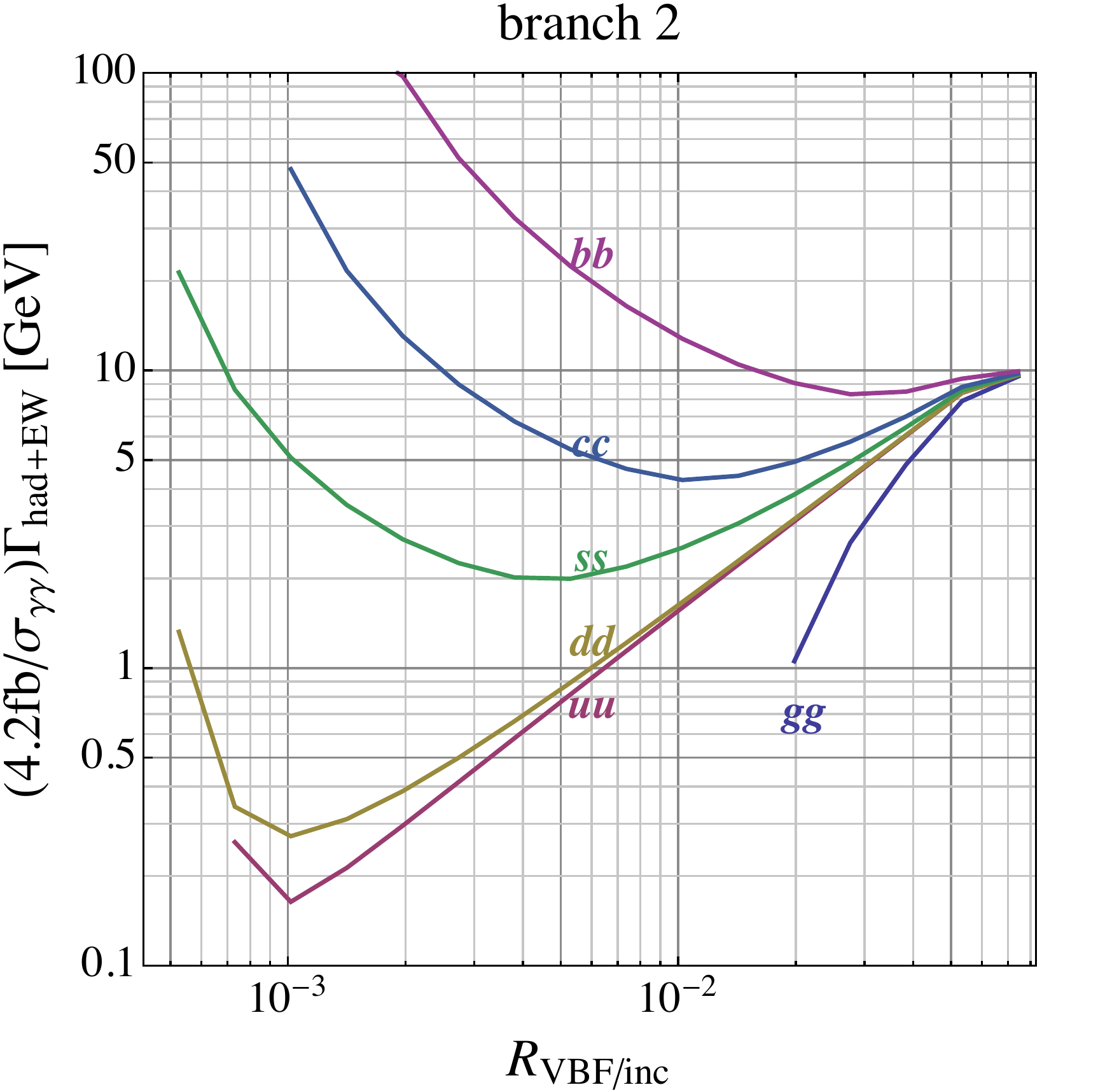} 
\caption{The maximum rate $(4.2 \, \, \text{fb} / \sigma_{\gamma \gamma}) \Gamma_\text{had+EW}$, which is a lower bound on the total rate $\Gamma_S$, for the spin-$0$ resonance $S$ decaying to hadronic final states and electroweak gauge boson as a function of the VBF/inclusive ratio $R_{\rm VBF/inc}$.  To find the maximum rate, we marginalize over the $\kappa$ parameter and consider both branch 1 (left) and branch 2 (right) for the ratios $\Rrd_{WW,ZZ}$, as a function of $\Rrd_{Z \gamma}$.  We allow $S$ to couple to electroweak gauge bosons and either $gg$, $b\bar b$, $c \bar c$, $s \bar s$, $d \bar d$, or $u \bar u$, as shown.   
}
\label{fig:width}
\end{figure}

The production rate of the resonance times the branching ratio to photons, $\sigma_{\gamma \gamma} = \sigma_{\rm inc}^{13\,{\rm TeV}} \times \BR_{\gamma \gamma}$, is fixed by the CMS and ATLAS observations, though the best-ft values of $\sigma_{\gamma \gamma}$ depend on the production channel (see Tab.~\ref{tab:Xsec}).
For spin-$0$, the inclusive cross section is given in~\eqref{eq:INCXsec0}.  The total decay rate $\Gamma_S$ may be written as 
\es{totalDecay}{
\Gamma_S &= \Gamma_{\gamma \gamma} \left(1  +  \Rrd_{Z\gamma} +  \Rrd_{ZZ} +  \Rrd_{WW} + \Rrd_{gg}+\Rrd_{u\bar{u}}+\Rrd_{d\bar{d}}+\Rrd_{s\bar{s}}+\Rrd_{c\bar{c}}+\Rrd_{b\bar{b}}  +\Rrd_{t\bar{t}} \right. \\
 &\left.+\Rrd_{e\bar{e}}+\Rrd_{\mu\bar{\mu}}+\Rrd_{\tau\bar{\tau}  }+\Rrd_\text{invisble} \right) \,.
}
Each of the ratios $\Rrd$ above is subject to constraints (see, for example,~\cite{Franceschini:2015kwy}).
We consider the possibility that the non-vanishing rates are those to electroweak gauge bosons and---as suggested by the current data---either $gg$, $b\bar b$, $c \bar c$, $s \bar s$, $d \bar d$, or $u \bar u$.
Current limits, scaled from 8\,TeV to 13\,TeV, then imply $\Rrd_{gg} \lsim 1300$ and $\Rrd_{b \bar b} \lsim 500$~\cite{Franceschini:2015kwy}.  As was shown in Sec.~\ref{sec:VBF}, the rates to gluons and quarks also determine the ratio $R^{0}_{\rm VBF/inc}$.  This allows us determine the width as a function of $R^{0}_{\rm VBF/inc}$ for these scenarios, as illustrated in Fig.~\ref{fig:width}.   

In Fig.~\ref{fig:width}, we show---for each channel---the maximum value of $(1 / \sigma_{\gamma \gamma}) \Gamma_\text{had+EW}$, where $\Gamma_\text{had+EW}$ is the rate to quarks (except top), gluons, and electroweak gauge bosons, as a function of the VBF ratio $R_{\rm VBF/inc}$.\footnote{Note that we do not include decays of $S$ to $hh$ pairs in $\Gamma_\text{had+EW}$ because this rate in principle depends on more parameters.  However, in general the rate to $hh$~\eqref{eq:Gammahh} is found to be small ($< 3$ GeV), except for a narrow range of $\Rrd_{Z \gamma} \in [0.5, 0.75]$, where---only for branch 1---it is possible for the rate $\Gamma_{hh}$ to be larger without violating current constraints. 
}
We normalize $\sigma_{\gamma \gamma}$ to $4.2$ fb. Note that to find the maximum we marginalize over the $\kappa$ parameter, given the current constraints.  The left (right) panel refers to branch 1 (branch 2) for the ratios $\Rrd_{ZZ, WW}$. Note that on the far right of these plots, the production is completely through EW gauge boson fusion.  At the other extreme, at very small values of $R_{\rm VBF/inc}$, the production is dominated by the heavy quarks or gluon fusion (depending on the relevant channel).  This is true for all channels except branch 1 of the gluon fusion channel, where both the smallest and largest values of $R_{\rm VBF/inc}$ are obtained through EW gauge boson fusion.  Interestingly, we find that that the maximum $\Gamma_\text{had+EW}$ is a strong function of $R_{\rm VBF/inc}$ for each channel.  Thus, dual measurements of $\Gamma_\text{had+EW}$ and $R_{\rm VBF/inc}$ may help illuminate the possibility that new light states are required in the EFT or that large branching ratios to currently-weakly-constrained final states, such as $t \bar t$, are required.   

The above discussion is relevant only for the spin-$0$ resonance; if the resonance is instead spin $2$, then the relationship between $R_{\rm VBF/inc}$ and the partial width to hadronic and electroweak states is not so clear.  In part, this is because---in this case---there is no strong separation in $R_{\rm VBF/inc}$ between quark anti-quark annihilation and vector boson fusion.  On the other hand, the partial width is in general much smaller in the spin-$2$ scenario since the production cross section is enhanced.  More specifically, after marginalizing over all production channels, we find that the constraint $(4.2 \, \, \text{fb} / \sigma_{\gamma \gamma} )\Gamma_\text{had+EW} < 5$ GeV.  Thus, if the width of the resonance is indeed $\sim$45 GeV and the resonance is spin $2$, then additional decays are required.  
Future data, combined with the formalism presented above, should help distinguish between these possibilities.

\section{Conclusions} \label{sec:Conclusions}

In this paper, we showed that the effective field theory of the new state $S$, combined with the standard model, leads to non-trivial signatures at the level of the LHC observables.  In Sec.~\ref{sec:correlations}, we concentrated on the decay of $S$ and showed that there are necessarily decays into other electroweak final states, with non-trivial relations between the branching ratios among the final states.  Importantly, the branching-ratio relations and sum rules we derived in that section will be tested in future runs of the collider.  In Sec.~\ref{sec:VBF} we asked a related question: given the effective field theory and the rate to diphotons observed so far, how can observables, such as ratios of rates that survive VBF cuts to inclusive rates, tell us about the coupling of $S$ to non-electroweak states, such as quarks.  We showed that the VBF ratio is a discriminator between gluon-fusion production and, for example, quark anti-quark production.  This is particularly relevant, given that in Sec.~\ref{sec:fit} we showed that current LHC data supports production through either gluon fusion or heavy quark annihilation.   Similarly, in Sec.~\ref{sec:SMrate} we showed that additional constraints on the coupling of the resonance to non-electroweak states are found when considering simultaneously the total width along with the VBF ratio.  In summary, we have provided a variety of simple relations directly among observables at the LHC which, in light of the low energy effective field theory, may have profound implications for understanding the nature of the excess as more data accumulates. 

\mysection{Acknowledgements}
We would like to thank Cedric Delaunay for collaboration in the early stages of this work. We also thank Liron Barak, Gilad Perez, Jesse Thaler for useful discussions.
J.F.K. acknowledges the financial support from the Slovenian Research Agency (research core funding No. P1-0035) and would like to the thank the CERN TH Department for hospitality while this work was being completed.
B.R.S. is supported by a Pappalardo Fellowship in Physics at MIT.  
The work of Y.S. is supported by the U.S. Department of Energy under grant Contract Number  DE-SC0012567.
J.Z. is supported in part by the U.S. National Science Foundation under CAREER Grant PHY-1151392.

\appendix

 \bibliography{paper_ref}

\end{document}